# Continuous Integration, Delivery and Deployment: A Systematic Review on Approaches, Tools, Challenges and Practices

Mojtaba Shahin[a], Muhammad Ali Babar[a], Liming Zhu[b]

[a] CREST – The Centre for Research on Engineering Software Technologies, The University of Adelaide, Australia
[b] Data61, Commonwealth Scientific and Industrial Research Organisation, Sydney, NSW 2015, Australia
mojtaba.shahin@adelaide.edu.au, ali.babar@adelaide.edu.au, liming.zhu@data61.csiro.au

***Abstract—Context***: Continuous practices, i.e., continuous integration, delivery, and deployment, are the software development industry practices that enable organizations to frequently and reliably release new features and products. With the increasing interest in and literature on continuous practices, it is important to systematically review and synthesize the approaches, tools, challenges, and practices reported for adopting and implementing continuous practices.

***Objective***: This research aimed at systematically reviewing the state of the art of continuous practices to classify approaches and tools, identify challenges and practices in this regard, and identify the gaps for future research.

***Method:*** We used systematic literature review (SLR) method for reviewing the peer-reviewed papers on continuous practices published between 2004 and 1st June 2016. We applied thematic analysis method for analysing the data extracted from reviewing 69 papers selected using predefined criteria.

***Results***: We have identified thirty approaches and associated tools, which facilitate the implementation of continuous practices in the following ways: (1) "reducing build and test time in continuous integration (CI)"; (2) "increasing visibility and awareness on build and test results in CI"; (3) "supporting (semi-) automated continuous testing"; (4) "detecting violations, flaws and faults in CI"; (5) "addressing security and scalability issues in deployment pipeline", and (6) "improving dependability and reliability of deployment process". We have also determined a list of critical factors such as "testing (effort and time)", "team awareness and transparency", "good design principles", "customer", "highly skilled and motivated team", "application domain", and "appropriate infrastructure" that should be carefully considered when introducing continuous practices in a given organization. The majority of the reviewed papers were validation (34.7%) and evaluation (36.2%) research types. This review also reveals that continuous practices have been successfully applied to both greenfield and maintenance projects.

***Conclusion***: Continuous practices have become an important area of software engineering research and practice. Whilst the reported approaches, tools, and practices are addressing a wide range of challenges, there are several challenges and gaps which require future research work for: improving the capturing and reporting of contextual information in the studies reporting different aspects of continuous practices; gaining a deep understanding of how software-intensive systems should be (re-) architected to support continuous practices; addressing the lack of knowledge and tools for engineering processes of designing and running secure deployment pipelines.

## I. INTRODUCTION

With increasing competition in software market, organizations pay significant attention and allocate resources to develop and deliver high-quality software at much accelerated pace [1]. Continuous Integration (CI), Continuous DElivery (CDE), and Continuous Deployment (CD), called continuous practices for this study, are some of the practices aimed at helping organisations to accelerate their development and delivery of software features without compromising quality [2]. Whilst CI advocates integrating work-in-progress multiple times per day, CDE and CD are about ability to quickly and reliably release values to customers by bringing automation support as much as possible [3, 4].

Continuous practices are expected to provide several benefits such as: (1) getting more and quick feedback from the software development process and customers; (2) having frequent and reliable releases, which lead to improved customer satisfaction and product quality; (3) through CD, the connection between development and operations teams is strengthened and manual tasks can be eliminated [5, 6]. A growing number of industrial cases indicate that the continuous practices are making inroad in software development industrial practices across various domains and sizes of organizations [5, 7, 8]. At the same time, adopting continuous practices is not a trivial task since organizational processes, practices, and tool may not be ready to support the highly complex and challenging nature of these practices.

Due to the growing importance of continuous practices, an increasing amount of literature describing approaches, tools, practices, and challenges has been published through diverse venues. An evidence for this trend is the existence of five secondary studies on CI, rapid release, CDE and CD [9-13]. These practices are highly correlated and intertwined, in which distinguishing these practices are sometimes hard and their meanings highly depends on how a given organization interprets and employs them [14]. Whilst CI is considered the first step towards adopting CDE practice [15], truly implementing CDE practice is necessary to support automatically and continuously deploying software to production or customer environments (i.e., CD practice). We noticed that there was no dedicated effort to systematically analyze and rigorously synthesize the literature on continuous practices in an integrated manner. By integrated manner we mean simultaneously investigating approaches, tools, challenges, and practices of CI, CDE, and CD, which aims to explore and understand the relationship between them and what steps should be followed to successfully and smoothly move from one practice to another. This study aimed at filling that gap by conducting a Systematic Literature Review (SLR) of the approaches, tools, challenges and practices for adopting and implementing continuous practices.

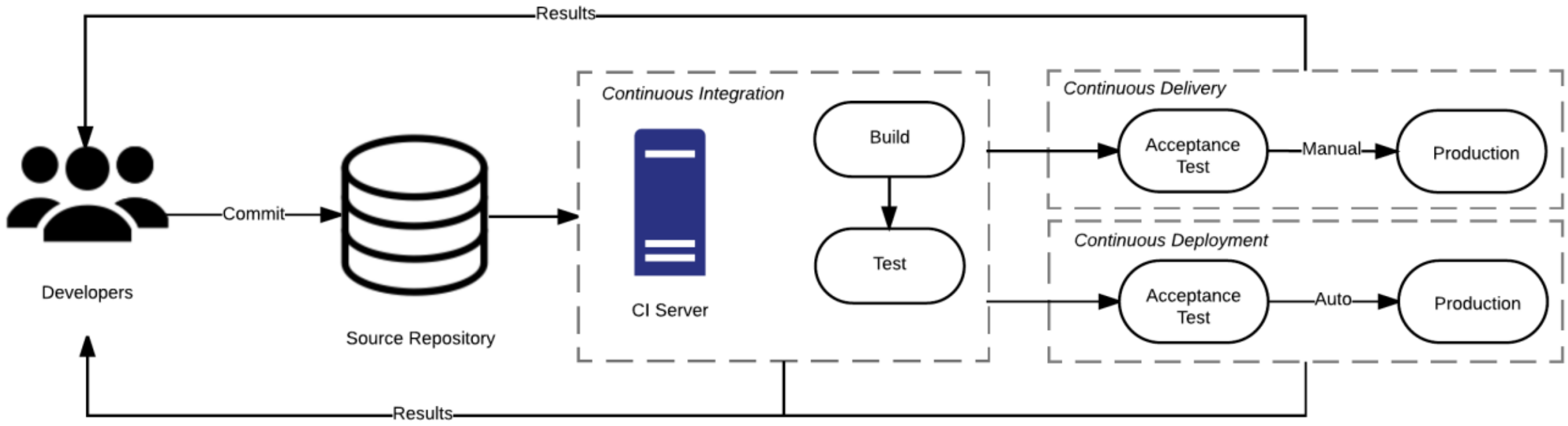

**FIGURE 1.** The relationship between continuous integration, delivery and deployment [16, 17]

This SLR provides an in-depth understanding of the challenges of adopting continuous practices and the strategies (e.g., tools) used to address the challenges. Such an understanding is expected to help identify the areas where methodological and tool support to be improved. This increases the efficacy of continuous practices for different types of organizations and software-intensive applications. Moreover, the findings are expected to be used as guidelines for practitioners to become more aware of the approaches, tools, challenges and implement appropriate practices that suit their industrial arrangements. For this review, we have systematically identified and rigorously reviewed 69 relevant papers and synthesized the data extracted from those papers in order to answer a set of research questions that motivated this review. The significant contributions of this work are:

1. A classification of the reported approaches, associated tools, and challenges and practices of continuous practices in an easily accessible format.
2. A list of critical factors that should be carefully considered when implementing continuous practices in both software development and customer organizations.
3. An evidence-based guide to select appropriate approaches, tools and practices based on the required suitability for different contexts.
4. A list of researchable issues to direct the future research efforts for advancing the state-of-the-art of continuous practices.

The rest of the paper is organized as follows: In Section II, we define continuous terminologies with summarizing related work and outlining the existing research gap. Section III describes the systematic literature review process with the review protocol. The quantitative and qualitative results of the research questions are described in Section IV. The Section V reports a discussion on findings. The threats to validity are discussed in Section VI. Finally, we present our conclusions in Section VII.

## II. FOUNDATIONS

### *A. Background*

Here we give an overview of continuous software engineering (e.g., continuous integration, continuous delivery, and continuous deployment) paradigm.

**Continuous software engineering** is an emerging area of research and practice. It refers to develop, deploy and get quick feedback from software and customer in a very rapid cycle [4, 18]. Continuous software engineering involves three phases: Business Strategy and Planning, Development and Operations. This study focuses on only three software development activities: continuous integration, continuous delivery and continuous deployment. Figure 1 shows the relationship between these concepts.

**Continuous Integration (CI)** is a widely established development practice in software development industry [4], in which members of a team integrate and merge development work (e.g., code) frequently, for example multiple times per day. CI enables software companies to have shorter and frequent release cycle, improve software quality, and increase their teams' productivity [4]. This practice includes automated software building and testing [5].

**Continuous DElivery (CDE)** is aimed at ensuring an application is always at production-ready state after successfully passing automated tests and quality checks [19, 20]. CDE employs a set of practices e.g., CI, and deployment automation to deliver software automatically to a production-like environment [15]. According to [6, 21], this practice offers several benefits such as reduced deployment risk, lower costs and getting user feedback faster. Figure 1 indicates that having continuous delivery practice requires continuous integration practice.

**Continuous Deployment (CD)** practice goes a step further and automatically and continuously deploys the application to production or customer environments [19, 22]. There is robust debate in academic and industrial circles about defining and distinguishing between continuous deployment and continuous delivery [4, 19, 20]. What differentiates continuous deployment from continuous delivery is a production environment (i.e., actual customers): the goal of continuous deployment practice is to automatically and steadily deploy every change into the production environment. It is important to note that CD practice implies CDE practice but the converse is not true [20]. Whilst the final deployment in CDE is a manual step, there should be no manual steps in CD, in which as soon as developers commit a change, the change is deployed to production through a deployment pipeline. CDE practice is a pull-based approach for which a business decides what and when to deploy; CD practice is a push-based approach [23]. In other words, the scope of CDE does not include frequent and automated release, and CD is consequently a continuation of CDE. Whilst CDE practice can be applied for all types of systems and organizations, CD practice may only be suitable for certain types of organizations or systems [20, 23, 24].

### *B. Existing literature reviews*

During this review, we also found five papers that have reported reviews on different aspects of continuous software engineering - two studies have investigated continuous integration in the literature [11, 13], two papers have explored continuous delivery [10] and deployment [9], and one study has targeted rapid release [12] (See Table 1). We summarize the key aspects of these studies. Stahl and Bosch [11] have presented a SLR on different attributes or characteristics of CI practice.

**TABLE 1.** Comparison of this SLR with existing secondary studies

| Study | Focus | # included papers | Search date |
|---|---|---|---|
| Stahl and Bosch [11] | CI | 46 | October 2012 |
| Eck et al. [13] | CI | 43 | N/A |
| Mäntylä et al. [12] | Rapid release | 24 | N/A |
| Rodríguez et al. [9] | CD | 50 | 27 June 2014 |
| Laukkanen et al. [10] | CI and CDE | 30 | February 2015 |
| This study | CI, CDE, and CD | 69 | 1 June 2016 |

That review has explored the disparity in implementations of CI practice in the literature. Based on 46 primary studies, the study had extracted 22 clusters of descriptive statements for implementing CI. The clusters have been classified into two groups: (a) *culled clusters* (e.g., fault frequency) which either came from one unique source or the literature interpreted and implemented them the same; and (b) *preserved clusters* (e.g., build duration) were described as statements that there is contention on them in the published literature. The paper proposed a descriptive model (i.e., the main contribution of the paper) to address the variation points in the preserved clusters.

Eck et al. [13] conducted a concept-centric literature review to study the organizational implications of continuous integration assimilation in 43 primary studies. The review revealed that organizations require implementing numerous changes when adopting CI. The study proposed a conceptual framework of 14 organizational implications (e.g., providing CI at project start) of continuous integration. The authors also conducted a case study of five software companies to understand the organizational implications of CI. Mäntylä et al. [12] performed a semi-systematic literature review to study benefits, enablers and problems of rapid release (including CI and CD) in 24 primary studies. The review did not comply with several of the mandatory aspects of a SLR's guidelines reported in [25] (e.g., lack of doing data extraction and analysis rigorously, including papers that were not found through search string). The review revealed that rapid releases are prevalent industrial practices that are utilized in several domains and software development paradigms (e.g., open source). It has been concluded that the evidence of the claimed advantages and disadvantages of rapid release is scarce. Rodríguez et al. [9] reported a systematic mapping study on continuous deployment to identify benefits and challenges related to CD and to understand the factors that define CD practice. Based on 50 primary studies, it has been revealed that moving towards CD necessitates significant changes in a given organization, for example, team mindsets, organization's way of working, and quality assurance activities are subject to change. The authors also found that not all customers are happy to receive new functionality on a continuous basis and applying CD in the context of embedded systems is a challenge. However, the main contribution of this mapping study lies in the identified 10 factors that define CD practice. For example, (a) fast and frequent release; (b) continuous testing and quality assurance; (c) CI; (d) deployment, delivery, and release processes and configuration of deployment environments.

We found that the work done by Laukkanen et al. [10] is the closest work to our study. They conducted a systematic review on 30 primary studies to identify the problems that hinder adopting CDE practice. The authors also reported the root causes for and solutions to the problems. The study grouped the problems and solutions into seven categories: build design, system design, integration, testing, release, human and organizational, and resource. The review [10] only focused on CDE practice rather than CD, in which the authors investigated CDE as a development practice where software is kept production-ready (i.e., CDE practice), but not necessarily deployed continuously and automatically (i.e., CD practice). Laukkanen et al. also revealed that the work of [9] used the term CD, while it actually referred to CDE practice. Furthermore, the SLR [10] indicated whilst it is interesting to study CD, but it was failed to find highly relevant literature on CD.

It is worth noting that it is common in software engineering to conduct several SLRs on a particular concept or phenomenon. To exemplify, there are four reviews (i.e., SLR or systematic mapping study) on technical debt [26]. What differentiates SLRs on a particular subject from each other is having different high level objectives, research questions, included studies and results. Having done a thorough analysis of the related reviews, we observed the following differences between this SLR and the existing reviews:

***Search string, inclusion and exclusion criteria***: Our search string, inclusion and exclusion criteria were significantly different with [9-13] for selecting the primary studies. Our work was aimed at reviewing papers that included empirical studies (e.g., case studies and experiments); we excluded the papers with less than 6 pages, which were included in [10, 11, 13]. It is important to note that the previous reviews except [10] used only automatic search, but we used both automated searches and snowballing for finding the relevant papers. Due to the aforementioned reasons, there is a significant difference in the papers reviewed by our SLR with the included papers in other SLRs. Out of 69 papers in our SLR, there were only 2, 10, 7, and 12 common papers with [9-11, 13] respectively.

***Research questions and results***: regarding ***RQ1*** and ***RQ2*** and their respective goals, there are no similar questions in other reviews. Both goals and results of ***RQ4*** are different to *RQ1* in [11, 13]. Whilst the objective of our research question (***RQ4***) was to comprehensively identify and analyze practices, guidelines, lessons learned and authors' shared experiences for successfully adopting and implementing each continuous practice, the given statements for implementing CI in [11] were not sufficiently abstracted and generalized and were not reported as practices for adopting and implementing CI. In fact, the main goal was to indicate there is a lack of consensus on implementing CI in practice. The focus of the review reported in [13] is on organizational aspects of assimilating CI practice rather than individual software projects. Furthermore, for both reviews [11, 13], the main contributions are model, conceptual framework, and empirical study rather than systematically summarizing, analyzing, and classifying the literature on CI. It is worth noting that due to having different coding schemes, level of details and emergence of categories, it was not easy to make one-to-one comparison of the identified challenges and practices between our SLR and [10]. However, our study identified a more comprehensive list of challenges, practices, guidelines, lessons learned and authors' shared experiences. Our findings show that we only have 5 common practices with [10]. Regarding ***RQ3***, there is a partial overlap among our SLR and the *RQ4* and *RQ1* in [9, 10] respectively.

**TABLE 2**. Research questions of this SLR

| Research Question | Motivation |
|---|---|
| **RQ1.** What approaches and associated tools are available to support and facilitate continuous integration, delivery and deployment? | To gain a comprehensive understanding of approaches (e.g., methods, algorithms, frameworks, techniques) and associated tools to facilitate implementation of continuous practices and to develop a classification of the approaches and tools. |
| **RQ2.** Which tools have been employed to design and implement deployment pipelines (i.e., modern release pipeline)? | Deployment pipeline is significantly important to move towards continuous practices (in particular CD/CDE). The idea of this question is to understand how researchers form deployment pipelines and which tools are employed to implement the deployment pipelines. It should be noted that the tools identified in **RQ1** can be also covered by this question provided that they are integrated and implemented in the deployment pipeline. |
| **RQ3.** What challenges have been reported for adopting continuous practices? | There might be obstacles and conflicts when adopting and implementing continuous practices in software provider and customer organizations. So, the idea with this question is to get an overview of different types of technical and organizational challenges, problems and constrains that the organizations might experience in transition to continuous practices. |
| **RQ4.** What practices have been reported to successfully implement continuous practices? | To identify good practices, guidelines, lessons learned and shared experiences when adopting and implementing CI, CDE, and CD. |

However, the goal of the questions has some overlaps with together, but closely looking at the result from each study, it clearly indicates a complementary relationship between them. Some of the major differences in the identified challenges are *lack of awareness and transparency*, *general resistance to change*, *distributed organization*, *team dependencies*, *customer environment*, *dependencies with hardware and other (legacy) applications*, which were not reported in the previous reviews [9, 10].

***Analyzing CI, CDE, and CD practices in an integrated manner***: As discussed earlier, CI, CDE and CD practices are highly correlated and intertwined concepts, in which there is no consensus on the definitions of these practices [27]. In our understanding to obtain a clear understanding of the approaches, tools, challenges and practices, it is essential to broadly study and cover CI, CDE and CD practices across its different dimensions, such as approaches, tools, contextual factors, practices, and challenges simultaneously in an integrated manner.

### C. Motivation for this SLR on continuous practices

According to [4], continuous software engineering includes a number of continuous activities such as continuous integration, delivery and continuous deployment. It is asserted that CI is a foundation for CDE, in which implementing reliable and stable CI practice and environment should be the first and highest priority for a given organization to successfully adopt CDE practice. We have mentioned that CDE and CD practices are frequently confused together and used interchangeably in the literature and practitioners' blogs. It is sometime hard to distinguish these correlated and intertwined practices. The meanings of these practices highly depend on who uses them [14, 27]. Since the main objective of this study is to systematically collect, analyze and classify approaches, tools, challenges and practices of continuous practices, we believe these practices, particularly CDE and CD practices, should be investigated together. Analysing CI, CDE, and CD practices in an integrated manner provides an opportunity to understand what challenges prevent adopting each continuous practice, how they are related to each other, and what approaches, associated tools, and practices exist for supporting and facilitating each continuous practice. Furthermore, this helps software organizations to adopt continuous practices step by step and smoothly move from one practice to another. We could not find any systematic review, which has studied these intertwined practices (i.e., integration, delivery, and deployment) together. The abovementioned reasons indicate the need of conducting a literature review tailored to the scope of the continuous integration, delivery and deployment in an integrated manner.

## III. Research Method

We used Systematic Literature Review (SLR) that is one of the most widely used research methods in Evidence-Based Software Engineering (EBSE) [28]. SLR aims at providing a well-defined process for identifying, evaluating, and interpreting all available evidence relevant to a particular research question or topic [25]. This research method involves three main phases: defining a review protocol, conducting a review, and reporting a review. Following the SLR guidelines reported in [25], our review protocol consisted of: (i) research questions, (ii) search strategy, (iii) inclusion and exclusion criteria, (iv) study selection, and (v) data extraction and synthesis. We discuss these steps in the following subsections:

### A. Research questions

This study aimed at summarizing the current research on "***continuous integration, continuous delivery and continuous deployment practices in software development***". We formulated a set of research questions (RQs) to be answered through this report. Table 2 summarizes the research questions as well as the motivations for them. The answers to these research questions can be directly linked to the objective of this SLR: an understanding of the available approaches and tools in the literature to support and facilitate CI, CDE, and CD practices (RQ1, RQ2), challenges (RQ3) and practices (RQ4) reported by empirical studies during adopting each continuous practices. The results of these research questions would enable researchers to identify the missing gaps in this area and practitioners to consider the evidence-based information about continuous practices before deciding their use in their respective contexts. It is worth noting that we distinguish between approaches and practices in this SLR. Cambridge and Longman dictionaries define *approach*, *method*, and *technique* similarly as the following "*a [special/planned/particular] way of doing something*"; however, *practice* is defined as "*the act of doing something regularly or repeatedly*" [29, 30]. In this SLR, we define *approach*, *method*, and *technique* as a technical and formalized approach to facilitate and support continuous practices [31]. For simplicity purpose, the approaches, methods, techniques, algorithms, and frameworks, along with the tools to support them, that are developed and reported in the literature for this purpose, are classified as *approach* rather than *practice*. On the other hand, software *practice* is a social practice [32] and is defined as shared norms and regulated rules and activities, which can be supported and improved by an approach [31, 33].

## B. Search strategy

In order to retrieve as many relevant studies as possible, we defined a search strategy [25, 34]. The search strategy used for this review is designed to consist of the following elements:

### 1) Search method

We used automatic search method to retrieve studies in six digital libraries (i.e., IEEE Xplore, ACM Digital Library, SpringerLink, Wiley Online Library, ScienceDirect, and Scopus) using the search terms introduced in Section III.B.2. We complemented the automatic search with snowballing technique [35].

### 2) Search terms

We formulated our search terms based on guidelines provided in [25]. The resulting search terms were composed of the synonyms and related terms about "continuous" AND "software". After running a series of pilot searches and verifying the inclusion of the papers that we were aware of, we utilized the final search string as presented in the following. It should be noted that the search terms were used to match with paper titles, keywords, and abstracts in the digital libraries (except SpringerLink) during the automatic search. The reason we included the "software" and its related terms in the search string was that continuous delivery and continuous deployment terminologies are also used in other disciplines (e.g., medicine). Therefore, we were able to avoid retrieving a large number of irrelevant papers.

***TITLE-ABS-KEY*** *(("continuous integration"* ***OR*** *"rapid integration"* ***OR*** *"fast integration"* ***OR*** *"quick integration"* ***OR*** *"frequent integration"* ***OR*** *"continuous delivery"* ***OR*** *"rapid delivery"* ***OR*** *"fast delivery"* ***OR*** *"quick delivery"* ***OR*** *"frequent delivery"* ***OR*** *"continuous deployment"* ***OR*** *"rapid deployment"* ***OR*** *"fast deployment"* ***OR*** *"quick deployment"* ***OR*** *"frequent deployment"* ***OR*** *"continuous release"* ***OR*** *"rapid release"* ***OR*** *"fast release"* ***OR*** *"quick release"* ***OR*** *"frequent release"* ***OR*** *"deployability"* ***OR*** *"continuous build"* ***OR*** *"rapid build"* ***OR*** *"fast build"* ***OR*** *"frequent build"* ***OR*** *"quick build")* ***AND*** *("software"* ***OR*** *"information system"* ***OR*** *"information technology"* ***OR*** *"cloud*"* ***OR*** *"service engineering"))*

### 3) Data sources

We queried six digital libraries, namely IEEE Xplore, ACM Digital Library, SpringerLink, Wiley Online Library, ScienceDirect, and Scopus for retrieving the relevant papers. According to [36], these are the primary sources of literature for potentially relevant studies on software and software engineering. For all these libraries, except SpringerLink, we ran our search terms based on title, keywords and abstract. It is important to note that currently SpringerLink search engine does not provide any facility for searching on the title, abstract and keywords [37]. We were forced to either restrict our search on the title only or apply search terms on the full text of the articles. While the former resulted in a quite few number of papers, the latter strategy returned more than 11700 papers. In order to address this situation, we followed the strategy adopted in [37]; we examined only first 1000 papers retrieved by search on the full text. However, we believe that Scopus was a complement to SpringerLink as Scopus indexes a large number of journals and conferences in software engineering and computer science [38, 39]. It is worth noting that Google Scholar was not selected as data source because of the low precision of search results and generating many irrelevant results [36].

## C. Inclusion and exclusion criteria

Table 3 presents the inclusion and exclusion criteria, which were applied to all studies retrieved from digital libraries. We did not choose a specific time as the starting point of the search period. Only peer-reviewed papers were included, and we excluded editorials, position papers, keynotes, reviews, tutorial summaries, panel discussions and non-English studies. Papers with less than 6 pages were excluded. We selected only those papers that have reported the approaches, tools, and practices using empirical research methods such as case study, experience report, and experiment. In cases where we found two papers addressing the same topic and have been published in different venues (e.g., in a conference and a journal), the less mature one was excluded. We eliminated duplicate studies retrieved from different digital libraries.

**TABLE 3**. Inclusion and exclusion criteria of this SLR

| **Inclusion Criteria** | |
|---|---|
| **I1** | A study that is peer-reviewed and available in full-text. |
| **I2** | A study that presents approaches (e.g., methods, techniques, frameworks, and algorithms) and associated tools to facilitate continuous practices or reports practices and challenges in adopting continuous practices. |
| **I3** | Empirical study: a study that evaluates, validates, or investigates the proposed approaches, tools and practices through empirical research methods such as case studies, survey, and experiments. |
| **Exclusion Criteria** | |
| **E1** | Non peer-reviewed papers such as editorials, position papers, keynotes, reviews, tutorial summaries, and panel discussions. |
| **E2** | Short papers (i.e., less than 6 pages). |
| **E3** | A study that is not written in English. |
| **E4** | Non-empirical studies (e.g., tool demo) |

## D. Study selection

Figure 2 shows the number of studies retrieved at each stage of this SLR. The inclusion and exclusion criteria were used to filter the papers in the following way:

**Phase 0:** We ran the search string on the six digital libraries and retrieved 14723 papers. Considering only first 1000 results from SpringerLink, we finally found 3942 potential papers.

**Phase 1**: We filtered the papers by reading title and keywords. When there were any doubts about the retrieved papers and it was not possible to determine the papers by reading the titles and keywords, these papers were transferred to the next round of selection for further investigation. At the end of this phase, 449 papers had been selected.

**Phase 2**: We looked at the abstracts and conclusions of the retrieved articles to ensure that all of them were related to the objective of our SLR. We applied snowballing technique [35] to scan the references of the selected papers in the second phase. We found 51 potentially relevant papers by title from the references of these 174 papers.

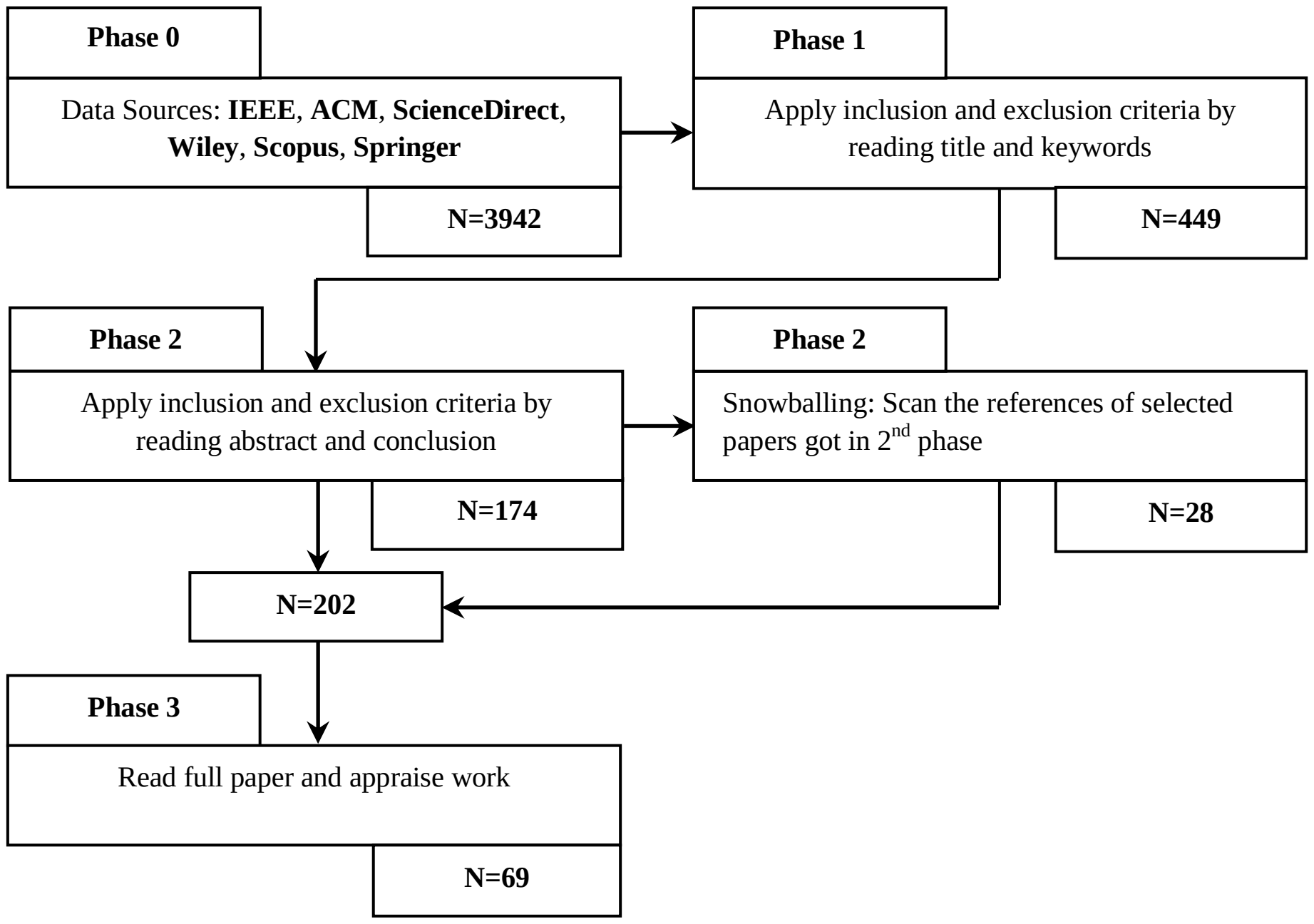

**FIGURE 2.** Phases of the search process

Inclusion and exclusion criteria were applied to the abstracts and conclusions of those 51 potentially relevant papers and we finally selected 28 papers for the next phase. It is important to mention that the main reason for conducting snowballing in this phase rather than applying it in the third phase, was to find as many relevant studies as possible.

**Phase 3**: In the last (third) selection round, we read the full text of the selected studies from second phase and if a paper met all the inclusion criteria, this paper was selected for inclusion in this SLR. We excluded the papers that were shorter than 6 pages, irrelevant, or whose full texts were not available. Furthermore, we critically examined the quality of primary studies to exclude those had low quality e.g., low reputation venues. We found four types of papers on continuous practices:

- Papers that present approaches (e.g., methods, techniques, frameworks, and algorithms) and associated tools to facilitate each continuous practice (RQ1).
- The second group consists of experience report papers which either present the challenges, problems, and confounding factors in adopting and implementing continuous practices (RQ3) or discusses practices, guidelines and lessons learned for this purpose (RQ4).
- A group of papers reporting surveys of the usage and importance of agile practices (e.g., continuous integration and delivery) in software development organizations.
- The papers in forth group used the concepts of continuous integration, delivery and deployment on developing and deploying an application, for example, applying CI practice on robotic systems, and mostly reported the potential benefits obtained by these concepts.

Since most papers in third and fourth groups did not meet any of research questions and were out of the objectives of this review, we excluded a large number of the papers in those groups. Finally, we selected 69 papers for this review. A significant part of the study selection, data extraction and synthesis phases has been conducted by first author. In each phase, we recorded the reasons of inclusion or exclusion decision for each of the papers, which were used for further discussion with second and third authors and reassessment whether a paper had to be included or not. A cross-check using a random number of the selected papers for each step was performed by the second author.

### *E. Data extraction and synthesis*

#### *1) Data extraction*

We extracted the relevant information from the selected papers based on the data items presented in Appendix B in order to answer the research questions of this SLR. It shows the research question(s) (described in Section III.A) that were supposed to be answered using different pieces of the extracted data. The extracted information was stored in MS Excel Spreadsheet for further analysis.

#### *2) Synthesis*

We divided the data extraction form into a) demographic and contextual attributes, b) approaches, tools, challenges, practices and critical factors of continuous practices. We used descriptive statistics to analyze the data items D1 to D10. In order to identify the research types (i.e., data item D7) reported in the selected papers, we classified them into six well-known research types: validation research, evaluation research, solution proposal, philosophical paper, opinion paper, and experience report [40]. The second set of data items (i.e., D11, D12, D13 and D14) were analyzed using qualitative analysis method, namely thematic analysis [41]. We followed the five steps of the thematic analysis method [41] as detailed below:

(1) Familiarizing with data: we tried to read and examine the extracted data items, e.g., D11 (approaches and tools), D12 (challenges), D13 (practices) and D14 (critical factors) to form the initial ideas for analysis.

(2) Generating initial codes: in the second step we extracted the initial lists of challenges, practices and factors for each continuous practice. It should be noted that in some cases, we had to recheck the papers.

(3) Searching for themes: for each data item we tried to combine different initial codes generated from the second step into potential themes.

(4) Reviewing and refining themes: the challenges, practices and critical factors identified from third step were checked against each other to understand what themes had to be merged with others or dropped (e.g., lack of enough evidence).

(5) Defining and naming themes: through this step, we defined clear and concise names for each challenge, practice and critical factor.

## IV. Results

Following subsections report the results from analyzing and synthesizing the data extracted from the reviewed papers to answer the research questions. The results are based on synthesizing the data directly collected from the reviewed papers with our minimal interpretations. We interpret and reflect upon the results in the discussion section.

### *A. Demographic attributes*

This subsection reports the demographic and research design attributes information: studies distribution, research types, study context and data analysis type, and application domains and project types. All of the included papers are listed in Appendix A.

#### *1) Studies distribution*

It is argued that reporting demographic information on the types and venues of the reviewed papers on particular research topic is useful for new researchers who are interested in conducting research on that topic. Therefore, the demographic information is considered one of the important pieces of information in an SLR. Figure 3 summarizes how 69 primary papers are distributed along the years and the different types of venues. The selected papers were published from 2004 to 2016. Note that the review only covers the papers published before 1st June 2016. In spite of continuous practices, in particular continuous integration and delivery are considered as the main practices proposed by agile methodologies (e.g., eXtreme Programming) introduced in early 2000, we were unable to find many relevant papers to our SLR before 2010. We found a couple of papers that conducted surveys on the usage and importance of agile practices (e.g., continuous integration and delivery) in software development organizations before 2010, but those papers have been excluded as they did not report any approach, practice and challenge regarding CI and CDE. It is argued that CDE and CD practices have recently been known and studied in academia (i.e., last 5 years) [42]. Figure 3 indicates a steady upward trend in the number of papers on continuous practices in the last decade. We noticed that 39 papers (56.5%) were published during the last 3 years, suggesting that researchers and practitioners are paying more attention to continuous practices. It is clear from Figure 3 that conference was the most popular publication type with 48 papers (i.e., 69.5%), followed by journal (14 papers, 20.2%), while only 7 papers [S15, S23, S28, S62, S63, S64, S65] came from workshops.

There are 11 out of 14 journal papers that have been published in 2015 and 2016, which indicates that the research in the area is becoming mature. Table 4 summarizes that the reviewed papers were published in 47 venues, in which *IEEE Software* and *International Conference on Agile Software Development (XP)* are the leading venues for publishing work on continuous practices research as they have published 10.1% (7 papers) and 8.6% (6 papers) of the reviewed papers. The *International Conference on Software Engineering* (i.e., 5 papers) and Agile Conference (e.g., 4 papers) maintained the subsequent positions. There are two venues (i.e., ITNG and RCoSE) with only two papers each. We note that more than half of the papers (40 out of 69, 57.9%) were published in 40 different venues. Some of the publication venues are not directly related to software engineering topics such as Robotic; it indicates that the research on continuous practices is being adopted by researchers in several areas that require software development.

**TABLE 4.** Distribution of the selected studies on publication venues

| Pub. Venue | # | % |
|---|---|---|
| IEEE Software | 7 | 10.1 |
| International Conference on Agile Software Development (XP) | 6 | 8.6 |
| International Conference on Software Engineering (ICSE) | 5 | 7.2 |
| Agile Conference | 4 | 5.7 |
| Information and Software Technology (IST) | 3 | 4.3 |
| International Workshop on Rapid Continuous Software Engineering (RCoSE) | 2 | 2.8 |
| International Conference on Information Technology: New Generations (ITNG) | 2 | 2.8 |
| Others | 40 | 57.9 |

#### *2) Research types*

This section summarizes the results from analyzing the data item D7 about research types. Table 5 shows that a large majority (49 out of 69, 70.9%) of the papers were reporting evaluation or validation research, in which they each correspond to 36.2% (25 papers) and 34.7% (24 papers) of the selected papers respectively. The high percentage of the evaluation research was not surprising because a noticeable number of the reviewed papers investigated and extracted challenges and practices of CI, CDE, and CD in industry through case studies with interview as data collection method (e.g., [S4]). That is why a vast majority of the papers in this category had used qualitative research approaches. Since prominent research methods of the validation papers are simulation, experiments, and mathematical analysis [40], 22 out of 25 papers in this category employed quantitative research methods. We also categorized 15 (21.7%) papers as personal experience papers, in which practitioners had reported their experiences from introducing and implementing one of the continuous practices. Solution proposal (5 papers) maintained the subsequent position. To give an example, [S9] collected opinions of three release engineers through interviews on continuous delivery's benefits and limitations, the required job skills, and the required changes in education. The reviewed papers were not fallen in the philosophical and opinion categories because we only included empirical studies.

#### *3) Study context and data analysis type*

We classified the reviewed papers into industry and academic cases. The industrial studies were carried out with industry or used real-world software-intensive systems to validate the proposed approach and tool; whilst academic category refers to those studies, which were performed in an academic setting. Our review reveals that a large majority of the reviewed papers (64 out of 69, 92.7%) are situated in the industry category, whilst only 6 [S1, S2, S16, S20, S22, S40] papers were conducted in academic settings.

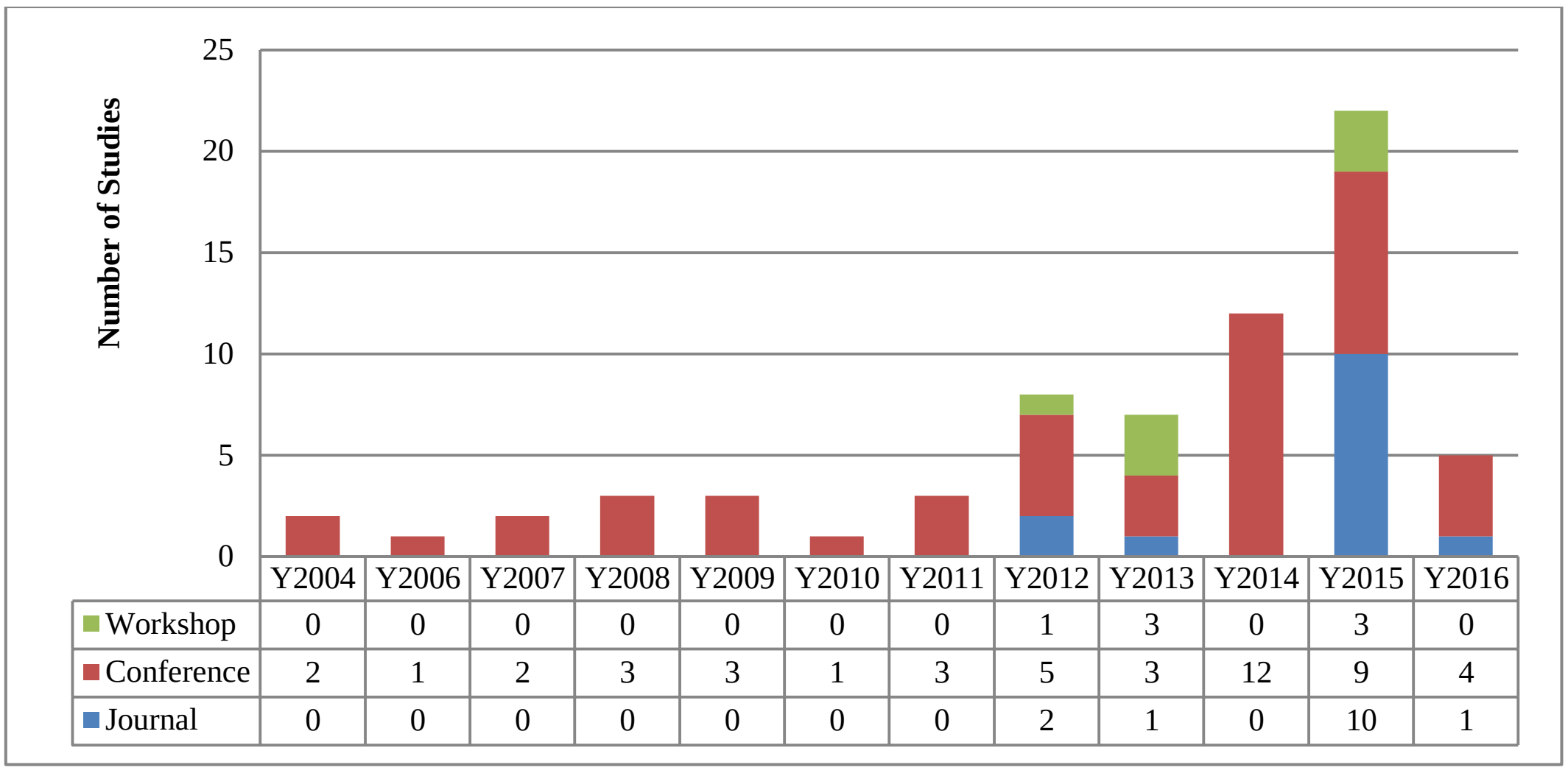

**FIGURE 3.** Number of selected studies published per year and their distribution over types of venues

**TABLE 5.** Number and percentage of papers associated to each research type and data analysis type

| | | **Data Analysis Type** | | | | |
|---|---|---|---|---|---|---|
| | | Qualitative | Quantitative | Mixed | Unclear | Total |
| **Research Type** | Evaluation Research | S5, S6, S7, S9, S10, S12, S13, S31, S36, S43, S45, S46, S56, S60, S62, S63 **(16)** | S18, S28, S51 **(3)** | S4, S11, S33, S41, S44 **(5)** | S30 **(1)** | **25 (36.2%)** |
| | Validation Research | S22, S67 **(2)** | S1, S2, S3, S8, S21, S23, S24, S27, S32, S34, S38, S40, S53, S54, S55, S61, S69 **(17)** | S16, S20, S25, S29, S64 **(5)** | **(0)** | **24 (34.7%)** |
| | Experience Report | S26, S37, S42, S49, S50, S52, S65 **(7)** | S39, S48 **(2)** | S14, S17, S57, S58 **(4)** | S15, S47 **(2)** | **15 (21.7%)** |
| | Solution Proposal | S35 **(1)** | S19, S59, S66, 68 **(4)** | **(0)** | **(0)** | **5 (7.2%)** |
| | Opinion Paper | **(0)** | **(0)** | **(0)** | **(0)** | **0 (0%)** |
| | Philosophical Paper | **(0)** | **(0)** | **(0)** | **(0)** | **0 (0%)** |
| | Total | **26 (37.6%)** | **26 (37.6%)** | **14 (20.2%)** | **3 (4.3%)** | |

It shall be noted that one paper [S40] has been placed into both categories as it conducted two case studies in academic and industry. The high percentage of the industry papers indicates a significant level of relevance and practicality of the results reported in this SLR. According to Table 5, there were the same number of reviewed papers that used qualitative and quantitative (26 out of 69, 37.6% each) research approaches, whilst we found 14 papers (20.2%), which employed both qualitative and quantitative research approaches for data analysis. It was not possible for us to specify data analysis method of three studies [S15, S30, S47] based on the provided information.

### *4) Application domains and project types*

We analyzed the data items D9 and D10 in Appendix B in order to provide potentially useful information for practitioners who are interested in project types and the domain specific aspects of the approaches, tools, challenges and practices reported for CI, CDE, and CD. Table 6 shows the application domains in which the reviewed approaches, practices and challenges can be placed. Regarding the application domain, not all the reviewed papers provided this information, which resulted in categorizing 38 studies under "unclear" category. For those papers that reported the application domains, we classified them into 13 application domains. The approaches, tools and practices introduced in one study can be applied in more than one application domains with several cases; for example, the continuous integration testing approach reported in [S40] has been applied in two different domains such as communication software and information management system. If one study uses more than one system as a case study, then we count this study N (number of systems) times in Table 6.

**TABLE 6.** Distribution of application domains of the selected studies

| Application domain | No. of Cases | Cases |
|---|---|---|
| Unclear | 39 | S1, S4, S5, S6, S9, S10, S12, S13, S16, S17, S18, S22, S23, S27, S28, S31, S33, S37, S38, S42, S44, S46, S47, S48, S49, S50, S55, S57, S58, S60, S61, S62, S63, S64, S65, S66, S67, S68 |
| Software/Web development framework | 13 | S3(2), S21, S24(4), S25, S32, S34, S36, S59(2) |
| Utility software | 9 | S3(3), S24(3), S30, S34(2) |
| Data management software | 8 | S3(2), S24(4), S25, S43 |
| Financial Software | 5 | S7(2), S41, S43, S45 |
| General software library | 5 | S2, S3(2), S24, S25 |
| Embedded system | 5 | S11, S19, S26, S52, S56 |
| Information management system | 4 | S8, S24, S40, S53 |
| Web server | 3 | S3(2), S24 |
| Communication software | 3 | S3, S24, S40 |
| Military Software | 3 | S11(2), S39 |
| Distributed system | 2 | S14, S54 |
| Web browser | 2 | S29, S69 |
| Other domains | 11 | S3 (4), S7, S15, S20, S51, S24, S34, S35 |

The work reported in [S34] uses two utility software as case studies, and x represents the number of cases in S34(x). It becomes clear from Table 6 that the "software/web development framework" domain has gained the most attention for continuous practices, followed by "utility software" and "data management software". We investigated the type of project (i.e., greenfield and maintenance) that continuous practices have been applied to. Our analysis of the data item D10 revealed that the greenfield and maintenance projects were reported in 17 and 16 papers respectively. However, there are 36 papers without any information about the types of projects for which the proposed continuous approaches, tools and practices had been applied.

### B. ***RQ1***. *What approaches and associated tools are available to support and facilitate continuous integration, delivery and deployment?*

We found 29 papers (42%) that reported approaches and associated tools to support and facilitate continuous integration, delivery or deployment practices. Table 7 lists all approaches and associated tools presented in the reviewed papers. The *Description* column provides a summary of the proposed approaches and associated tools. Third column indicates the proposed approaches and tools have been mainly used and applied to facilitate what continuous practices. We classified the available approaches and associated tools into six groups depending on their features and/or the areas in which they were used as the followings. Apparently, the six categories are not mutually exclusive, as there were several approaches and tools fallen in more than one category. For brevity purpose, we only elaborate a small subset of the studies as examples.

#### *1) Reduce build and test time in CI*

The approaches and tools in this category aim at minimizing the total time in the build process and test phase, which consequently improves performance and efficiency of continuous integration practice [S3, S19, S23, S25, S34, S55, S64, S67]. Since slow build process can be an obstacle to practicing continuous integration, Bell et al. [S3] proposed two approaches namely VMVM (Virtual Machine in a Virtual Machine) and VMVMVM (Virtual Machine in a Virtual Machine on a Virtual Machine) to isolate in-memory and external dependencies among test cases respectively. Whilst eliminating in-memory dependencies between tests enables running each test in its own process, which significantly reduces the overhead of dependencies among short test cases, VMVMVM approach executes the long-running test cases in parallel. The combination of VMVM and VMVMVM accelerates the total build time, which can relieve a deployment pipeline from long-running builds.

A number of papers [S34, S55, S64] in this category developed approaches that reduce the time of test execution by selecting a set of tests cases and prioritizing them, in which developers are enabled to receive the results early in the testing process. To give an example, Elbaum et al. [S55] proposed CRTS (Continuous Regression Test Selection) and CTSP (Continuous Test Suite Prioritization) approaches to effectively run regression tests within continuous integration development environments. The proposed approaches use test suite execution history data to improve the cost-effectiveness of pre-submit testing (i.e., tests performed by developers before committing code to repository) and reduce test case execution costs.

McIntosh et al. [S25] revealed that in C and C++ applications there might be header files that not only increase the time of rebuild process, but also due to frequent maintenance requires significant effort. Thus, these header files, called hotspots, are bottleneck to continuous integration build process. Through analysis of the Build Dependency Graph (BDG) and the change history of a system, the proposed approach in [S25] enables team to identify the header files that should be optimized first to improve build performance. Hence, the team members only can focus on header files with added value.

### 2) Increase visibility and awareness on build and test results in CI

As the frequency of code integration increases, the information (build and test results) produced during practicing CI would increase exponentially. This may considerably slow down the feedback in CI. Therefore, it is critical to collect and represent the information in timely manner to help stakeholders to gain better and easier understanding and interpretation of the results. Several studies [S1, S2, S13, S22, S24, S33, S38, S52, S64, S67] have reported approaches and associated tools for improving developers' understanding of their projects' status when implementing CI practice. The authors of [S2] found that stand-alone CI tools (e.g., Jenkins) produce huge amount of data that may not be easily utilized by stakeholders (e.g., developers and testers). They reported a framework and platform called SQA-Mashup to integrate and visualize the information produced in CI-toolchain using two views: (1) dynamic view, which is a visualization view for developers and testers and (2) time view, which indicates a chorological view on events (i.e., failure event) happened in CI-toolchain. It was found that interpretation of the proposed views is time-consuming and should be performed by professionals (e.g., tester). Brandtner et al. [S24] proposed a rule-based approach, named SQA-Profile, to classify stakeholders based on their activities in CI environment. The project-independent SQA-Profile enables tailoring and dynamic composition of scattered data in CI system. Nilsson et al. [S13] have found that companies need to describe and arrange testing activities and efforts before moving to CI. CIViT (Continuous Integration Visualization Technique) aims at visualizing end-to-end process of testing activities. CIViT enables team members to avoid duplicate testing efforts and visually understand the status (i.e., time and extent) of testing of quality attributes.

### 3) Support (semi-) automated continuous testing

There are 7 papers that have proposed approaches and tools for (semi-) automating tests in deployment pipelines [S19, S32, S38, S40, S52, S53, S54]. Two papers [S40, S53] have provided frameworks to support Continuous Integration Testing (CIT) in SOA systems. Whilst the work reported in [S40] partly automates test case generation in CIT using sequence diagrams as input, Surrogate, the simulation framework proposed by [S53], enables CIT for partial implementation. Through this framework, bugs can be identified when some components or even all components are still unavailable. Kim et al. [S38] proposed NHN Test Automation Framework (NAFT) as an integrator for existing CI servers to facilitate CI practices through automating repetitive and error-prone processes for testing. It aids communication among various stakeholders using tables to represent tests and test environments.

### 4) Detect violations, flaws and faults in CI

Addressing the failures and violations in continuous integration systems, particularly at the early stage of development are the targets of several papers [S16, S21, S25, S32, S33, S34, S42, S52, S53, S54, S55]. For example, one study [S16] reported an approach and associated tool called WECODE to automatically and continuously detect software merge conflicts earlier than a version control system is used by developers. The tool enables developers to detect the conflicts in uncommitted code that version control systems are not able to detect. In [S21], the authors developed a method includes incremental integration with simple and true backtracking in order to reduce the impacts of broken builds in the context of component-based software development. In the normal situation, a failure in the build process of a component stops the integration process. The failure should be resolved and the component needs to be rebuilt. But the incremental integration method addresses this issue by building components using earlier build results of the same components. This approach leads the integration process becomes more resilient against build failures.

### 5) Address security and scalability issues in deployment pipeline

Our literature review has identified only two papers dealing with security issue in deployment pipelines [S27, S66]. Gruhn et al. argued that continuous integration systems are vulnerable for security attacks and misconfiguration [S27]. Having proposed a secure build server, they encapsulated build jobs using virtualization environment with snapshot capability to prevent one project's security attacks from infecting other projects' build jobs in multitenant CI systems. In [S66], it has been discussed that the security of a deployment pipeline may be threatened by malicious code being deployed through the pipeline and direct communication between components in the testing and production environments. Rimba et al. [S66] proposed an approach, which integrates security design fragments (i.e., security patterns) through four compassion primitives namely connect tactic, disconnect tactic, create tactic, and delete tactic to secure deployment pipelines. For a large-scale software project, the full build can take hours as it includes compilation, unit testing and acceptance testing. Roberts [S47] has extended normal continuous integration process and proposed Enterprise Continuous Integration (ECI) approach to split up project into several modules using binary dependencies. Despite every module has its own CI, ECI provides the feedback that single-project CI provides. ECI addresses scalability issue in normal CI and enables small teams continuously integrate with binary dependencies developed by other teams.

### 6) Improve dependability and reliability of deployment process

Some papers [S8, S59, S68] dealt with deployment process of applications that have adopted continuous delivery or deployment practices. The work reported in [S8] investigated the reliability issue in high-frequency releases of Cloud applications. It has been argued that two major contributing factors i.e., cloud-infrastructure APIs (EC2 API) and deployment-tool (i.e., OpsWorks[1] and Chef[2]) can affect the reliability of cloud applications when they adopt continuous delivery and deployment. Four error-handling approaches have been implemented on rolling upgrade tool to deal with reliability issues and facilitating continuous delivery. Increasing the frequency of deployment (e.g., by adopting CD practice) would make error diagnosis harder during sporadic operations [S68]. An approach, called Process Oriented Dependability (POD), has been proposed to improve dependability of deployment process in cloud-based systems. The POD approach models the sporadic operations as processes through collecting metrics and logs in order to alleviate the difficulty of error diagnosis in deploying cloud-based systems on a continuous basis.

[1] https://aws.amazon.com/opsworks/
[2] https://www.chef.io/chef/

**TABLE 7.** A classification of approaches and associated tools to facilitate continuous integration, delivery and deployment: ❶ (reduce the build and test time in CI); ❷ (increase visibility and awareness of build and test results in CI); ❸ (support (semi-) automated continuous testing); ❹ (detect violations, faults, and flaws in CI); ❺ (address security and scalability issues in deployment pipeline); ❻ (improve dependability and reliability of deployment process)

| Description of Approaches and Tools | Category | Apply to |
|---|---|---|
| **Wallboard technique** [S1]: It indicates the current integration and delivery status of all branches within a project. | ❷ | All |
| **SQA-Mashup** [S2]: It can integrate and visualize data produced in CI environments. | ❷ | CI |
| **VMVM/VMVMVM** [S3]: It is used to isolate in-memory and external dependencies among test cases. | ❶ | CI |
| **Error-handling approaches on rolling upgrade** [S8]: A set of error-handling approaches to deal with reliability issues which are inherent to cloud environments. | ❻ | CDE/CD |
| **CIViT** [S13]: It is used to visualize end-to-end process of testing activities in transformation to continuous integration. By visualizing end-to-end process of testing activities, team members are enabled to reduce testing efforts. | ❷ | CI |
| **WECODE** [S16]: It automatically and continually detects software merge conflicts earlier than a version control system is used by developers as well as detects conflicts in uncommitted code. | ❹ | CI |
| **uBuild** [S19]: It provides continuous testing making reproducible and deterministic tests in order to achieve automated build. | ❶❸ | CI |
| **Backtracking Incremental Continuous Integration** [S21]: Through simple and true backtracking approaches, this approach increases the resilience of build process against failures and ensures that a working version is available at all times. | ❹ | CI |
| **BuildBot Robot** [S22]: It notifies who is responsible for test failure in CI environment in friendly and funny way. It makes continuous integration environment visible to all developers. | ❷ | CI |
| **Hydra** [S23]: It is Nix-based continuous build tool, which automatically produces build environment for projects. Therefore, it can reduce the efforts to maintain a continuous integration environment. | ❶ | CI |
| **SQA-Profile** [S24]: Through a set of rules, it can provide a dynamic composition of CI dashboards based on stakeholder activities in tools of a CI environment. | ❷ | CI |
| **Hotspot Approach** [S25]: In order to have fast build system in continuous integration infrastructure, this approach identifies header files that are bottlenecks for build process. | ❶❹ | CI |
| **Secure Build Server** [S27]: It extends the default build server in CI environment using encapsulating infected build jobs and prevents spreading infection to other build jobs in multitenant CI systems. | ❺ | CI |
| **Automatic and agile testing of product lines based on combinational interaction testing** [S32]: It makes automatic testing as an integrated part of continuous integration framework and enables developers of software product lines to identify potential interaction faults in build process. | ❸❹ | CI |
| **Ambient awareness based approach** [S33]: It enhances build status awareness among team members, which results in decreasing the number of broken builds and a strong sense of responsibility towards failures in build process. | ❷❹ | CI |
| **Integrating fault localization and test case prioritization technique in CI** [S34]: The fault location and test case prioritization approaches are combined to support commit built in continuous integration, which consequently improves efficiency (time) and effectiveness of the whole CI process. | ❶❹ | CI |
| **NHN Test Automation Framework** [S38]: It supports CI practices through automating repetitive and error-prone processes for testing in a continuous integration environment. It aids communication among various stakeholders using tables to represent tests and test environments. | ❷❸ | CI |
| **Continuous Integration Testing for SOA** [S40]: A Unified Test Framework (UTF) for Continuous Integration Testing (CIT) of SOA, which would partly automate test case generation in CIT using sequence diagrams as input. | ❸ | CI |
| **User-defined Script** [S42]: It supports enforcement at commit time by establishing a pre-commit step (i.e., Subversion pre-commit hook) to force developers into fixing the violation in code commit. It does not allow committing codes that is not compliable with conventions (e.g., as-build architecture). | ❹ | CI |
| **Enterprise Continuous Integration** [S47]: A modified version of normal continuous integration process to split up the project into several modules using binary dependencies. Despite every module has its own CI, ECI provides the feedback that single-project CI provides. | ❺ | CI |
| **Tinderbox** [S52]: It is a continuous integration and automated testing system, which helps find integration problems earlier in development cycle, reduce the cost to fix the integration problems and improve visibility and awareness among team members. | ❷❸❹ | CI |
| **Surrogate** [S53]: A simulation framework to implement continuous integration testing for SOA systems when some components or even all components are still unavailable. With this, bugs are identified at the early stage of | ❸❹ | CI |
| development. | | |
| **CiCUTS** [S54]: It integrates CUTS, system modeling executing tool, with continuous integration environments. Through this approach, developers and testers are enabled to continuously perform system integration testing on target environments using emulation approaches and identify performance problems before components are completely implemented. | ❸❹ | CI |
| **Continuous Regression Test Selection (CRTS)** [S55]: It enables running effectively regression testing within continuous integration development environments. The technique improves the cost-effectiveness of pre-submit testing (i.e., tests performed by developers before committing code to repository) and reduce test case execution costs. | ❶ | CI |
| **Continuous Test Suite Prioritization (CTSP)** [S55]: It can reduce delays in fault detection during post-submit testing (i.e., all tests that are performed after code submitted to repository). In overall, it can improve the cost-effectiveness of the continuous integration process. | ❶❹ | CI |
| **Rondo** [S59]: Adopting continuous deployment in dynamic environments such as pervasive computing environments is associated with a number of challenges. Deployment process in such environments should be reproduced in different sites, support customizability, and should be equipped with custom rollback mechanism. Rondo, an automation tool, satisfies all above-mentioned requirements to facilitate adopting continuous deployment practice in dynamic environments. | ❻ | CDE/CD |
| **Code-Churn Based Test Selection (CCTS)** [S64]: This technique analyses code churns and test execution results to select an optimal subset of test suites on system level. So, it helps large-scale software development organizations speed up CI. It enables team members to gain a better understanding of number of test failures. | ❶❷ | CI |
| **Enhancing the security design of a deployment pipeline** [S66]: This approach integrates four security design fragments (i.e., security patterns) at the design level to secure deployment pipelines. The security of the pipeline is ensured through not allowing malicious code is deployed through the pipeline and preventing direct communication between components in the testing and production environments. | ❺ | CDE/CD |
| **Morpheus** [S67]: It facilitates CI practice through improving the quality of feedback, in which each developer only receives the test results of his own changed code (i.e., easy interpretation of test results). Additionally, in order to minimize build and test time, it executes automated tests in the environment that is similar to the production environment. | ❶❷ | CI |
| **Process Oriented Dependability (POD)** [S68]: An approach to improve dependability of deployment process in cloud-based systems. This approach models the sporadic operations as processes in order to alleviate the difficulty of error diagnosis during sporadic operations when CD practice is adopted and implemented. | ❻ | CDE/CD |

### *C. **RQ2**. Which tools have been employed to design and implement deployment pipeline?*

This section presents the findings to answer to RQ2. Deploying software on a continuous basis to end users has increased the importance of deployment pipelines [42]; the success of adopting continuous practices in enterprises heavily relies on deployment pipelines [1]. Hence, the choice of appropriate tools and infrastructures to make up such pipeline can also help mitigate some of the challenges in adopting and implementing continuous integration, delivery and deployment practices. We have investigated the deployment toolchain reported in the literature and the tools for implementing deployment pipelines. Since continuous delivery and deployment might be used interchangeably, we used the term deployment pipeline, which is equal to the modern release engineering pipeline [42], instead of continuous integration infrastructure, or continuous delivery or deployment pipeline.

A deployment pipeline should include explicit stages (e.g., build and packaging) to transfer code from code repository to the production environment [1, 43]. Automation is a critical practice in deployment pipeline; however, sometime manual tasks (e.g., quality assurance tasks) are unavoidable in the pipeline. It is worth noting that there is no standard or single pipeline [1]. Our literature reveals that only 25 out of 69 studies (36.2%) discussed how different tools were integrated to implement toolchain to effectively adopt continuous practices. It should be noted that the tools reported in this section are mostly existing open sources and commercial tools, which aim to form and implement a deployment pipeline. However, the tools discussed in Section IV.B are intended to facilitate the implementation of continuous practices. These tools can be also used as part of deployment pipeline implementation provided that they are integrated and evaluated in the pipeline. As shown in Figure 4, we divided the deployment pipeline into 7 stages: (i) version control system; (ii) code management and analysis tool; (iii) build tool; (v) continuous integration server; (vi) testing tool; (vii) configuration and provisioning; and (viii) continuous delivery or deployment server. It should be noted that not all stages are compulsory as well as we could not find any primary study among the 25 studies that had implemented a pipeline involving all stages mentioned in Figure 4. At the first stage, developers continually push code to code repository. The most popular version control systems used in deployment pipelines are *Subversion*[3] and *Git/GitHub*[4] as each has been reported in 6 papers. We found 7 papers [S2, S14, S18, S20, S42, S52, S62], which used code management and analysis tools as part of deployment pipeline to augment build process. The work reported in [S20] integrated *SonarQube*[5] into *Jenkins*[6] CI server for gathering metric data such as test code coverage and coding standard violations and visualized them to developers. Continuous integration servers check the code repository for changes and use automated build tool [44].

[3] https://subversion.apache.org/
[4] https://github.com/git/git
[5] www.sonarqube.org/
[6] https://jenkins-ci.org/

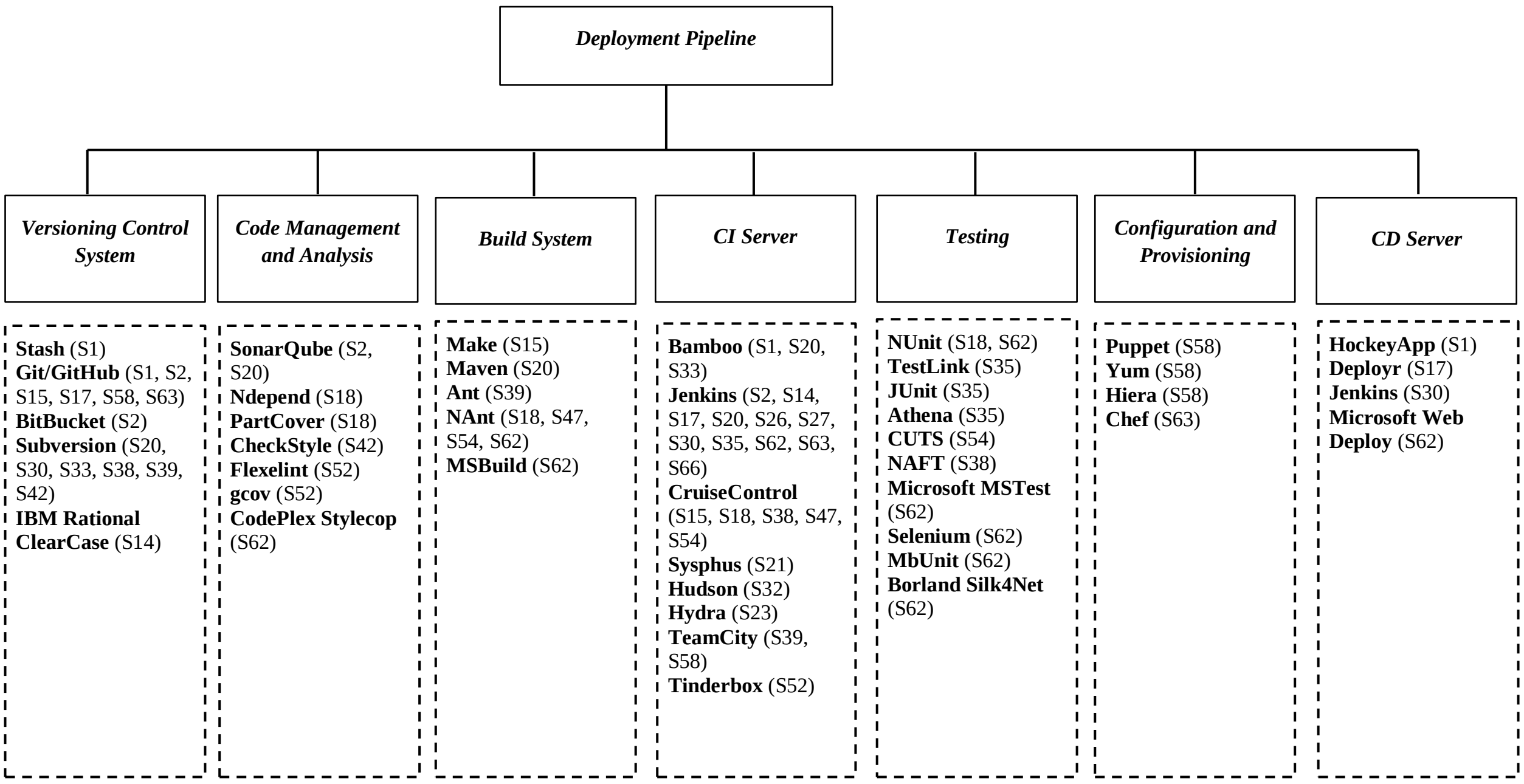

**FIGURE 4.** An overview of tools used to form deployment pipeline

Through CI servers, it is possible to automatically trigger build process and run unit tests. *Jenkins* [S2, S14, S17, S20, S26, S27, S30, S35, S62, S63, S66] has gained the most attention among existing CI servers in the literature. It should be noted that some CI servers (e.g., *Jenkins*, *Bamboo*[7] and *Hudson*[8]) are also able to deploy software to staging or production environment [45]. A study reported in [S30] used *Jenkins* as continuous delivery/deployment server.

*Bamboo* and *CruiseControl* maintained the subsequent positions. [S39, S58] used TeamCity as CI server in the pipeline and other CI servers have been reported in one paper each. The next step of deployment pipeline is to run a set of tests in various environments. There are only four papers [S18, S35, S54, S62], which integrated testing tools as part of deployment pipeline. Two papers [S35, S18] employed *JUnit*[9] and *NUnit*[10] for unit test in the pipeline respectively, while one paper [S35] also used a test runner called *Athena* to execute test suites and store the results in a format that can be used by Jenkins. Furthermore, *TestLink*[11] as a test management framework has been employed to store the results of acceptance tests run in different sites. The work reported in [S54] combined *CUTS* as a system modeling executing tool to *CruiseControl*[12] to enable developers and testers to continuously run system integration tests at the early stages of the software lifecycle (i.e., before complete system integration time) of component-based distributed real-time and embedded systems. The tool can capture performance metrics of executing systems such as execution time, throughput, and the number of received events. It is asserted that providing automated configuration of servers and virtual machines is one of innovations in deployment pipelines [42]. That can be the reason why we observed only two studies [S58, S63] that used configuration management tools as integrated part of deployment pipeline to streamline the configuration and provisioning tasks. One study [S1] used *HockeyApp*[13] as continuous delivery server to distinguish external release from internal one as well as it enables to deliver a build as a release to customers. The cases reported in [S17, S62] respectively used a Ruby-based software deployment called *deoloyr* and *Web Deploy* tool to automatically deploy code to production.

### D. ***RQ3****. What challenges have been reported for adopting continuous practices?*

This section summarizes the results of RQ3, *"What challenges have been reported for adopting continuous practices?"* As discussed in Section III.E.2, we analyzed the data item D12 using the thematic analysis method [41] for identifying and synthesizing the challenges for moving to and adopting CI, CDE, and CD. Our analysis resulted in the identification of 20 challenges, which are shown in Table 8. We provide detailed descriptions of the identified challenges as a follow:

#### *1) Common challenges for adopting CI, CDE and CD practices*

Under this category, we list the challenges of implementing all continuous integration, delivery and deployment practices together. Most of the challenges are usually associated with introducing any new technologies or phenomena in a given organization.

#### *a) Team Awareness and Communication*

***Lack of awareness and transparency*:** Our review has identified several papers that report a lack of sufficient awareness among team members may break down transition towards continuous practices [S6, S10, S31, S43, S45, S50, S56, S62]. Espinosa et al. [46] defined "awareness" as short-term knowledge about a team and its tasks. Continuous delivery process should be designed in a way that the status of a project, number of errors, the quality of features, and the time when features are finished are visible and transparent for all team members [S10, S31, S43, S50]. The work reported in [S31] asserted a lack of sufficient knowledge about the changes made in the main branch during developing work

[7] https://www.atlassian.com/software/bamboo/
[8] hudson-ci.org/
[9] junit.org/
[10] http://www.nunit.org/
[11] testlink.org/
[12] http://cruisecontrol.sourceforge.net/
[13] http://hockeyapp.net/features/

packages by self-organized teams resulted in increased number of merge conflicts in delivery.

***Coordination and collaboration challenges*****:** Some of the reviewed studies also reported that successfully implementing continuous practices requires more collaborations and coordination between all team members [S4, S6, S10, S41, S45, S56, S62]. For example, compared to less frequent release, deploying software on a continuous basis requires more communication to and coordination with operations teams [S4]. Gmeiner et al. [S62] argued that real benefits of deployment pipeline can be obtained by having a common understanding and collective responsibilities among all stakeholders. Another study [S41] noted that there is a need of strong coordination and communication between release manager and other team members (e.g., testers) to improve the release process. Laukkanen el al. [S45] reported that coordination and collaboration challenges as result of adopting continuous integration in distributed teams.

### *b) Lack of Investment*

***Cost*****:** Cost and investment play an important role in embracing continuous practices in both customer and software development organizations. Several of the reviewed studies [S4, S6, S12, S27, S37, S43, S45, S49, S57, S62] reported that practicing efficiently each of the continuous integration, continuous delivery or deployment is associated with high cost that can be attributed to many factors. For example, a study [S37] reported that a major resource upgrade was needed to support CI practice. Gruhn et al. [S27] observed that adopting continuous integration in Free, Libre and Open Source Software (FLOSS) requires extra computation, bandwidth, and memory resources. CI systems are required to perform build jobs, which include downloading patch files, compiling new versions of code, and running a large set of unit and acceptance tests. The work reported in [S43] revealed that performing automated acceptance tests in the deployment pipeline requires a significant amount of resources from customers. Two studies [S57, S62] observed that building, improving, and maintaining infrastructures (e.g., deployment pipeline) for continuous deployment practice needed a significant amount of time, money and training. There was also cost associated with training and coaching team members to adopt continuous practices [S57].

***Lack of expertise and skill*****:** Several papers [S4, S5, S6, S12, S45, S49, S57] reported a significant gap in the required skills when implementing continuous practices. This is mainly because most of the practices (e.g., test and deployment automation) associated with CI, CDE, and CD demand new technical and soft (e.g., communication and coordination) skills and qualifications. Several studies [S4, S6, S57] indicated the needs of highly skilled developers for practicing CD.

***More pressure and workload for team members*****:** It has been reported that building high-quality applications that are supposed to be frequently released to customers may cause some team members to face more stress and extra efforts [S4, S5, S6, S45, S49, S58]. Callanan and Spillane [S58] discussed that operations team was under more pressure to deliver software on a continuous basis. The study reported in [S49] has found that transforming a six month release into continuous release noticeability increased the workload of the developers and the release team. Whilst the transition forced developers to more analyze their codes in order to thoroughly identify negative side effects of their codes, the release team experienced difficulties to find issues in release process. One reason for this pressure could be that team members are directly responsible for affecting their customers' experiences.

***Lack of suitable tools and technologies***: According to eleven studies [S5, S6, S8, S10, S27, S43, S49, S56, S57, S60, S66], the limitations of existing tools and technologies are inhibitors to achieving the goals of continuous practices. Researchers pointed out [S5, S10] that the existing tools are inefficient in reviewing code and providing feedbacks from test activities in continuous integration. They emphasized that test automation is not sufficiently provided by current infrastructure. Other studies [S8, S27] highlighted the build and deployment tools employed in the deployment pipeline are vulnerable to security and reliability issues. Analysing the reliability issue in high-frequency releases of Cloud applications revealed that using external resources and cloud-based tools in a deployment pipeline leads to increased errors and delays, which consequently hinders continuous delivery practice [S8]. Olsson et al. [S10] indicated that the high frequency changes in tools and the need of learning new tools are the major barriers to adjust to continuous integration. Three papers [S56, S57, S60] revealed that the current tools and technologies either have limited functionalities or cannot enable all organization to truly adopting CD practice. To exemplify, a study [S56] reported that lack of appropriate technologies hindered automatically and continuously deploying applications in embedded system domain with customer-specific environments.

### *c) Change Resistance*

***General resistance to change*****:** Whilst employees generally resist to change, people may embrace changes provided that there are convincing reasons for those changes [47]. Introducing continuous practices may necessitate adopting a new way of working for some team members (e.g., accepting more responsibilities by developers). The reviewed studies reported that objections to change were a barrier to move towards and successfully implement continuous practices [S4, S5, S6, S12, S56, S57, S62]. A study [S62] found that establishing the necessary mindset required by a continuous delivery was a time-consuming process; another study [S5] concluded that changing the old habits of developers was problematic when introducing CI. Our investigation revealed that the team members were unwilling to change their ways of working due to lack of trust and rapport on the benefits of continuous practices, fear of exposing low quality code, and suffering more stresses and pressures.

***Scepticism and distrust on continuous practices*****:** Six papers [S4, S5, S6, S12, S45, S49] referred to lack of trust and scepticism about the added values that may bring by adopting continuous practices as potential risks for moving towards these practices. To give an example, the experience reported in [S49] revealed that the release team was worried about allowing several concurrent releases. This is mainly because continuous release might bring side effects for them and make them unable to identify which release was causing which problem. In addition, another study [S12] reported that lack of trust in application's quality may reduce the confidence of team members to move from CI to CD and deploy the application to production on a continuous basis.

**TABLE 8**. A classification of challenges in adopting CI, CDE, and CD practices

| | | Challenges | Key Points and Included Papers | # |
|---|---|---|---|---|
| Common Challenges for Adopting CI, CDE, and CD | Team Awareness and Communication | **Ch1**. Lack of awareness and transparency | ▪ Lack of awareness and transparency in the delivery process [S6, S10, S31, S43, S45, S50, S56, S62]<br>▪ Lack of understanding about the status of project increased number of merge conflicts [S31] | 8 |
| | | **Ch2**. Coordination and collaboration challenges | ▪ Practicing CI, CDE, CD needs more and effective coordination and communication between team members [S4, S6, S10, S41, S45, S56, S62] | 7 |
| | Lack of Investment | **Ch3**. Cost | ▪ Major upgrade in infrastructures and resources [S4, S6, S12, S27, S37, S43, S45, S49]<br>▪ Training and coaching [S57, S62] | 10 |
| | | **Ch4**. Lack of experience and skill | ▪ CI, CDE, and CD demand new technical and soft skills [S4, S5, S6, S12, S45, S49, S57]<br>▪ Need highly skilled developers [S4, S6, S57] | 7 |
| | | **Ch5**. More pressure and workload for team members | ▪ More stress for developers and operations team [S4, S5, S6, S45, S49, S58]<br>▪ More responsibilities for developers [S49] | 6 |
| | | **Ch6**. Lack of suitable tools and technologies | ▪ Lack of mature tools for automating tests and reviewing code in CI [S5, S6, S10, S43, S49]<br>▪ Frequency changes in tools [S10]<br>▪ Security and reliability issues in build and deployment tools [S8, S27, S66]<br>▪ Current tools don't fit to all organizations [S56, S57, S60] | 11 |
| | Change resistance | **Ch7**. General resistance to change | ▪ Changing the old habits of team members [S4, S5, S6, S12, S56, S57, S62]<br>▪ Time-consuming process to change team mindset [S62] | 7 |
| | | **Ch8**. Scepticism and distrust on continuous practices | ▪ Lack of trust on benefits of CI, CDE, CD [S4, S5, S6, S12, S45, S49] | 6 |
| | Organizational processes, structure and policies | **Ch9**. Difficulty to change established organizational polices and cultures | [S4, S6, S10, S12, S43]<br>▪ Lack of agile and suitable business model [S10, S12]<br>▪ Changing long-lived feature branching to short-lived one in an established company [S43] | 5 |
| | | **Ch10**. Distributed organization | ▪ Distributed team model [S12, S37, S45]<br>▪ Inconsistent perceptions among team members [S12, S45] | 3 |
| Challenges for Adopting CI | Testing | **Ch11**. Lack of proper test strategy | ▪ Lack of fully automated testing [S4, S5, S12, S36, S41, S43, S45]<br>▪ Lack of test-driven development [S12] | 7 |
| | | **Ch12**. Poor test quality | ▪ Instable tests [S4, S5, S6, S41, S45, S50, S62]<br>▪ Low test coverage [S56]<br>▪ Low quality test data [S6]<br>▪ Long running tests [S4, S5, S45, S50]<br>▪ Test dependencies [S5, S41] | 8 |
| | Merging conflicts | **Ch13.** Merging conflicts | [S4, S6, S21, S31, S41, S45]<br>▪ Third party components [S45]<br>▪ Incompatibly among dependent components [S31]<br>▪ Lack of understanding about changed components [S31] | 5 |
| Challenges for Adopting CDE | Lack of suitable architecture | **Ch14.** Dependencies in design and code | [S4, S5, S6, S10, S31, S41, S57, S60]<br>▪ Highly coupled architectures [S60]<br>▪ Difficulty to find autonomous requirements for frequent integrations [S5] | 8 |
| | | **Ch15.** Database schemas changes | ▪ Frequent changes in database schema [S6, S57, S58, S62] | 4 |
| | Team dependencies | **Ch16**. Team dependencies | ▪ Cross-team dependencies [S6, S31, S45, S50, S56, S57]<br>▪ Ripple effects of changes on multiple teams [S50]<br>▪ Dependency between feature team and module team in embedded system domain [S56] | 6 |
| Challenges for Adopting CDE | Customer challenges | **Ch17**. Customer environment | ▪ Lack of access to customer environment [S56, S60]<br>▪ Complex and manual configuration [S10, S62]<br>▪ Diversity and complexity of customer sites [S4, S6, S10, S29, S43]<br>▪ Difficulty to stimulate production-like environment [S56, S60] | 8 |
| | | **Ch18**. Dependencies with hardware and other (legacy) applications | ▪ Releasing an application on continuous basis requires deploying all dependent applications in customer site [S6, S10, S29, S43, S56, S62]<br>▪ Hardware and network dependencies [S56] | 6 |
| | | **Ch19.** Customer preference | ▪ Not all customers happy with frequent release [S6, S29, S43]<br>▪ Customer organization policy may affect practicing CD [S57] | 4 |
| | Domain constrains | **Ch20**. Domain constrains | ▪ Some domains don't allow or cause difficulties to truly adopt and implement CD [S4, S5, S6, S9, S10, S24, S31, S41, S44, S48, S56, S57, S60, S65] | 14 |

### *d) Organizational Processes, Structure and Policies*

***Difficulty to change established organizational polices and cultures*****:** According to [48], the organizational culture is a set of habits, behaviours, attitudes, values and management practices adopted by an organization. Two studies [S10, S12] discussed the difficulties in changing organizational cultures for aligning with the principles of continuous practices. Based on a study, Olsson et al. [S10] reported that being traditionally a hardware-oriented company was an obstacle in transition towards CI practices, however, [S12] highlighted this issue as the case company used to have six month release cycle. Both papers revealed lack of suitable and agile business model in organizations resulted in negative consequences for continuous practices. Rissanen and Münch [S43] found that practicing the short-lived feature branching, which is regarded as one of the best practices in continuous delivery is not easy to apply in a company with long established practices.

***Distributed organization*****:** It has been reported that practicing continuous integration and deployment in distributed development teams can be associated with a number of challenges (i.e., lack of visibility) [S12, S37, S45]. In both cases [S12, S45], the authors argued that introducing CI practice in distributed development model was challenging. That is mainly because it would prohibit having consistent perceptions among distributed teams and decrease the visibility of development sites. In an experience reported by Sutherland and Frohman [S37], it has been asserted that the distributed development model adopted by Scrum team was a barrier to CI practice. It is mainly because allocating a dedicated and private integration server environment to each individual Scrum team led to detecting integration issues that have been postponed to a very large extent. As a result, the team was forced to put all teams onto a single server environment.

### *2) Challenges for adopting CI practice*

#### *a) Testing*

***Lack of proper test strategy:*** One of the most prominent roadblocks to adopting continuous integration reported by several studies was the challenges associated with testing phase. Whilst it is asserted that automated test is one of the most important parts of successfully implementing CI, the case organizations studied [S4, S5, S12, S36, S41, S43, S45] were unable to automate all types of tests. Lack of fully automated testing may stem from different reasons such as poor infrastructure for automating tests [S12], time-consuming and laborious process for automating manual tests [S43] and dependencies between hardware and software [S5]. Whilst lack of test-driven development (TDD) practice has been reported in [S12] as a barrier to establishing CI practice, Debbiche et al. [S5] have revealed that regardless of TDD being practiced or not, a huge dependency between code and its corresponding tests made integration step very complicated. The work reported in [S36] revealed that although automating Graphic User Interface (GUI) testing through applying a set of GUI testing tools could partially alleviate the challenges of rapid release, but due to reliability concerns, the quality assurance (QA) members were needed to manually check the system during running automatic test.

***Poor test quality:*** The next challenge in testing phase during CI adoption is about low test quality. This includes having unreliable tests (i.e., frequent test failures) [S4, S5, S6, S41, S45, S50, S62], high number of test cases [S50], low test coverage [S56] and long running tests [S4, S5, S45, S50]. These issues not only can impede the deployment pipeline, but also can reduce the confidence of software development organizations to automatically deploy software on a continuous basis. Rogers [S50] observed that the number of tests grows in large-codebase and they run slowly. Therefore, developers are not able to receive the feedback from tests quickly and practicing CI starts to break down. To give another example, the author of [S62] found that it is hard to stabilize tests at the user interface level.

#### *b) Merging Conflicts*

Our review has revealed that conflicts during code integration causes bottlenecks for practicing CI [S4, S6, S21, S31, S41, S45]. There are several reasons for these conflicts that can occur when integrating code: one study [S45] reported that third-party components caused severe difficulty to practice CI. Sekitoleko et al. [S31] observed that incompatibility among dependent components and lack of knowledge about changed components caused teams facing extra effort to rewrite their solutions. It is asserted that merge conflicts are mainly attributed to highly coupled design [S31, S41].

### 3) Challenges for adopting CDE practice

#### a) Lack of Suitable Architecture

We found several studies discussing that unstable application architectures create hurdles in smooth transition towards continuous delivery and deployment practices.

***Dependencies in design and code*:** Some authors [S4, S5, S6, S10, S31, S41, S57, S60] asserted that inappropriately handling dependencies between components and code cause challenges in adopting continuous integration and in particular continuous delivery and deployment practices. The work reported in [S10] argued that the existence of huge dependency between components and the dependency between components interfaces resulted in highly dependent development teams and ripple effect of changes. It has been concluded that highly coupled architectures can cause severe challenge for CDE practice because changes are spanned across multiple teams with poor communications between them [S57, S60]. There was only one paper [S5], which considered software requirements as a challenge for CI as the interviewees reported that (i) finding the right size of requirements for being tested separately when broken down is challenging; (ii) it is not easy to understand whether small changes that do not directly add value to a feature are worth integrating or not.

***Database schemas changes***: Technical problem relating to database schemas changes should be effectively managed in the deployment pipeline. A few number of the reviewed studies [S6, S57, S58, S62] revealed that frequent changes in database schema as a technical problem when moving to continuous delivery. One study [S6] in this category highlighted that small changes in code resulted in constant changes in database schemas. Another study [S62] argued that a large part of concern in configuration of the automated test environment involved setting up databases. The study reported in [S57] discussed that one of the studied case companies did not put extra effort to streamline its database schema changes, which resulted in severe bottlenecks in its deployment process.

#### b) Team Dependencies

Team structures and interactions among multiple teams working on a same codebase system play an important role in successfully implementing CDE and CD practices. Several of the reviewed studies [S6, S31, S45, S50, S56, S57] reported that high cross-team dependencies prohibited development teams to develop, evolve and deploy applications or components and services into production independently of each other. This issue also has major impact on practicing CI as a small build break or test failure may have ripple effects on different teams [S50]. The author of [S56] argued that feature and module (hardware) teams developing embedded domain systems were highly dependent, in which each feature was complied, tested and built by a combination of both teams. This required a strong and proper communication and coordination among them. Two studies [S50, S57] in this group also discussed that nonexistence of a suitable architecture can increase the cross-team dependency.

### 4) Challenges for adopting CD practice

It has been noted that CD practice may not be suitable to any organizations or systems. We discuss the challenges and barriers that can limit or demotivate organizations from adopting CD practice.

#### a) Customer Challenges

***Customer environment***: A set of papers discussed that diversity and complexity of customers' sites [S4, S6, S10, S29, S43], manual configuration [S10, S62], and lack of access to customer environment [S56, S60] may cause challenges for team members when transferring software to customers through CD practice. According to [S4, S43], continuously releasing software product to multiple customers with diverse environments was quite difficult as it was needed to establish different deployment configurations for each customer's environment and component's version. A small set of papers [S56, S60] reported that it was not easy, if possible, to provide production-like test environment. Lwakatare et al. [S56] also observed that lack of access to and insufficient view on customer environment complicated simulating production environment. The aforementioned issues caused organizations challenges in providing fully automated provisioning and automated user acceptance test.

***Dependencies with hardware and other (legacy) applications***: Our analysis has revealed that albeit an application might be production-ready, dependencies between the application with other applications or hardware may be roadblocks to transition from CDE to CD practices (i.e., deploying the application on a continuous basis) [S6, S10, S29, S43, S56, S62]. It means it is needed to ensure that there is no integration problem when deploying an application to production. For example, a study [S10] reported that an increased number of upgrades and new features made the networks highly complex with the potential of becoming incompatible with legacy systems. The authors of [S56] found that dependency with hardware and compatibility with multiple versions as challenge for steady and automatically deploying software into customer environment.

***Customer preference:*** Some studies considered the preference of customers and their policies as important factors which should be carefully considered to move towards CD practice. It was revealed that not always customers are pleased with continuous release due to frequent update notifications, broken plug-in compatibility and increased bugs in software [S6, S29, S43]. Customer organization's policy and process may not allow truly implementing CD, as in an experience report Savor et al. [S57] reported that banks did not allow them to continuously push updates into their infrastructures.

#### b) Domain Constrains

A software system's domain is a significant factor that should be considered when adopting continuous deployment practice [S4, S5, S6, S9, S10, S24, S31, S41, S44, S48, S56, S57, S60, S65]. A large-scale qualitative study by Leppänen et al. [S4] indicated that domain constraints could change the frequency of deploying software to customers as well as the adoption of deployment method (e.g., calendar-based deployment). Compared with telecommunication and medical systems, web applications more frequently embrace the frequent deployment. In [S24], it has been reported that despite continuous integration practice was successfully adopted by a case company, it was not possible to fully apply continuous deployment practice on safety critical systems. We found two studies discussing the challenges of adopting CD in embedded systems [S56] and pervasive systems [S65].

### E. **RQ4**. What practices have been reported to successfully implement continuous practices?

This section reports the findings from analysis of the data extracted (i.e., D13) to answer RQ4, "*What practices have been reported to successfully implement continuous practices?*" Similar to RQ3, we first provide a high level classification of practices to understand which practices can be applied to each CI, CDE, CD and which practices are common for all CI, CDE, and CD. Table 9 presents 13 practices and lessons learnt reported in the reviewed papers.

#### 1) Common practices for implementing CI, CDE, and CD

##### a) Improve Team Awareness and Communication

In Section IV.B.2, we discussed how approaches and associated tools can increase a project's visibility and transparency for adopting continuous practices. This section reports the analysis of a few papers [S6, S31, S37, S43, S44, S47, S49] that provided practices for increasing team awareness and communication. Robert [S47] observed that appropriately labelling the latest version of client source and keep updating the server version in client-server application enabled developers to understand when everything is working together. In order to make changes visible for customer, a study [S44] in this category suggested recording the changed features in a change log to enable customers to track what and when features have changed. Marschall [S49] suggested that team members be regularly informed (e.g., by email) about branches that are completely out-dated. We found four papers [S6, S31, S37, S44] that argued that knowledge sharing practice should be consolidated among team members as enablers for adopting CI [S31, S37] and improvement for rapid release [S44].

##### b) Investment

***Planning and documentation***: It is argued that establishing continuous practices in a given organization necessitate planned and structured steps for clearly defining and documenting all the business goals and development activities [S28, S31, S36]. This is considered helpful to minimize the challenges associated with continuously releasing software features [S28, S31, S36]. Bellomo et al. [S28] observed that weaving requirements and designs through prototyping at the beginning of a release planning cycle enabled the studied team to smooth continuous delivery process. The release level prototyping with quality attributes focus enabled product owner and architect to work closely for quickly responding to prototype feedback. The case organization studied in [S58] developed a standard release path (i.e., a set of rules) for application packaging and deployment for which all the steps and activities to production are determined. This enabled the organization to easily embrace CD and release frequently and with confidence. Adopting CD should be slow with preparing, understanding and documenting engineering processes. For example, one of the case companies studied in [S57] spent 2 years to institutionalize CD practice. Five studies [S6, S11, S17, S37, S43] emphasized the importance of documentation when adopting continuous practices. It has been suggested that continuous activities (build, test, and packaging) should be well documented to help different stakeholders to understand the history of the activities in deployment pipeline. For example, Ståhl and Bosch [S11] proposed a descriptive *Integration Flow Model* for enabling team members to describe and record integration flow implementations in software development companies. The model consists of "input" (e.g., binary repository), "activity" (e.g., packaging) and "external triggering factors (e.g., scheduling)" elements.

***Promote team mindset***: As discussed earlier, lack of positive mindset about continuous practices is a confounding factor in adoption of these practices. Two papers [S5, S45] reported that organizational management organized CI events, which were run by the team who built the CI infrastructure to spread the positive mindset about CI. In order to encourage new developers to commit code several times per day, Facebook runs a six-week boot camp [S48] to help developers to overcome their fear of code failure. Another paper [S57] argued giving freedom to developers (e.g., full access to the company's code) enabled them to feel empowered to release new code within days of being hired.

***Improve team qualification and expertise***: Our review has identified the practices that aim at improving team qualification and expertise to bridge the skills gap to successfully implement continuous practices. We found several studies [S5, S6, S45, S48, S57] that provided formal training and coaching (for example through events) arranged by organizations. For instance, OANDA, a company studied in [S57], assigned new developers to the release engineering team for several months in order to get trained and familiar with CD practice. Claps et al. [S6] reported a software provider that leveraged CI developers' experience for transition from CI to CD by integrating automated continuous deployment of software into the existing CI workflow of developers to ensure there is no, or a low learning curve.

##### c) Clarifying Work Structures

Our analysis identified the practices that emphasize the importance of clarification of the work structures in successfully adopting and implementing continuous practices.

***Define new roles and teams***: A noticeable practice is defining new roles and responsibilities in software development lifecycle when a project adopts continuous practices [S1, S9, S29, S30, S45, S48, S49, S51]. Krusche and Alperowitz [S1] defined hierarchical roles such as release manager and release coordinator to introduce continuous delivery to multi-customer projects. Another work [S29] indicated that using a dedicated build sheriff role proved successful in practicing CI. The build sheriff engineer not only watches the build machine continuously but also aids developers by identifying and resolving the backouts that previously had to be addressed by developers. Another case [S45] reported the rotational policy implemented to enable team members to take different responsibilities to get higher understanding about the status of CI process. Another study [S57] also reported similar practice as developers were encouraged to rotate between different teams. Hsieh and Chen [S30] advocated having a *single responsible person* in team to constantly authorize and watch CI system. This helps to prevent ignoring broken builds by developers, particularly those happen during overnight. It was also reported that establishing a temporary or dedicated team to facilitate transitions towards continuous practices was helpful. The experience reported in [S37] highlighted that establishing a virtual Scrum team with expertise in infrastructures and operations was helpful to mitigate potential risks in software release. Another study [S5] observed the usage of pilot team who trained other team members and provided guidelines about CI goals to them through workshops and meetings to stimulate CI concepts. Two studies reported the establishment of a dedicated team for design and maintenance of infrastructure and deployment pipeline. This helps organizations in CD transformation [S57] and reduces release cycle time [S58].

***Adopt new rules and policies***: Several studies have reported the need of new rules, regulations, policies and strategies for enabling CI/CD [S26, S39, S45, S46, S48, S50, S58]. For example, one company [S39] enforced developers to solve the errors occurred during their commits in less than 30 minutes or revert the check-in. A paper [S46] reported a set of rules for improving deployability such as: creating tests cases at the system-level should take one day on average. In another paper [S26], the authors argued that having deployable software all the time has been reached by the following rule "*whenever a test complained, the integration of a change set failed, and the software engineer is obliged to update the test code or production code*".

### 2) Practices for implementing CI

This category presents three types of practices namely *improving testing activity*, *branching strategies* and *decomposing development into smaller units*, to enable and facilitate practicing CI.

#### a) Improve Testing Activity

Whilst Sections IV.B.1 and IV.B.3 summarized a set of approaches and tools proposed in the literature for improving test phase during CI, this section discusses three practices for this purpose. Karvonen et al. [S12] indicated that adopting test-driven development (TDD) and daily build practices are essential for CI practice. Neely and Stolt [S17] reported that one of the appropriate practices for removing manual tasks of QA was "***test planning***". This practice stimulates close collaboration between QA and developers to document a comprehensive list of automated tests. They argued that this practice liberates QAs from manually testing the majority of the software applications for regression bugs [S17]. The authors in [S39] suggested another practice called "***cross-team testing***", which means integration test of module *A* should be performed by programmers or testers who have not been involved in the implementation of module *A*. It has been argued that this practice helped detect more defects and build an objective appreciation of the modules. Rogers [S50] argued that the problem of slow unit tests in CI system can be alleviated by separating them from functional and acceptance tests.

#### b) Branching Strategies

Branching is a well-known CI practice. The practices such as repository use [S30, S44] and short-lived feature branching [S43] were presented as software development practices that support CI. Short-lived branching also supports the adoption of CDE practice as one study [S43] reported that an organization changed the long-lived feature branches to short-lived and small ones for exposing new features faster to the clients to receive feedback faster. Two studies [S29, S48] reported the practice of having developers to commit changes to a local repository and later on those changes would be committed to a central repository. However, in one case [S29], the code that passed all build and automated tests would be committed to the central repository by build sheriffs (i.e., introduced in Section IV.E.1.c). In this way, a release process will be more stable. It was also reported that having many branches hampers practicing CI. Feitelson et al. [S48] observed that working on a single stable branch of the code reduces time and effort on merging long-lived branches into trunks.

#### c) Decompose Development into Smaller Units

A set of the reviewed papers [S5, S10, S30, S36, S45, S47, S48, S49, S50, S51, S57] emphasized that software development process be decomposed into smaller units to successfully practice CI, but none of them provided concrete practice for this purpose. The main goal of this type of practice is to keep build and test time as much small as possible and receive faster feedback. Three papers [S10, S48, S49] argued that large features or changes should be decomposed into smaller and safer ones in order to shorten the build process so that the tests can be run faster and more frequently. For cross-platform applications, the complexity of dependency between components increases dramatically and it can be an obstacle to applying CI to them. Hsieh and Chen proposed a set of patterns namely *Interface Module*, *Platform Independent Module* and *Native Module* to control dependency between modules of cross-platform applications [S30]. They suggested that the platform-independent code should be placed into *Platform Independent Module* and these modules should be built in the local build environment. Through this pattern not only the build time reduces, but also the build scripts remain simple. Another paper [S5] proposed dead code practice, which can reduce dependency between components before integration through activating and testing a code or component only if all dependencies among them are in place. Decomposing development process into independent tasks enables organizations to have smaller and more independent teams (e.g., cross-functional teams), which was argued as an enabler for fully practicing CI [S50] and CDE [S51, S57].

### 3) Practices for implementing CDE

#### a) Flexible and Modular Architecture

As discussed in Section IV.D.3.a, technical dependency between codes or components can act as an obstacle to adopt CDE and CD. The reviewed studies reported that delivering software in days instead of months requires architectures that support CDE adoption [S7, S12, S28, S30, S45, S51, S57]. The software architecture should be designed in a way that software features can be developed and deployed independently. Loosely coupled architecture minimizes the impact of changes as well. For example, Laukkanen et al. [S45] observed that the studied organization had to re-architect their product (e.g., removing components caused trouble) to better adopt CI and CDE. It is also asserted that teams that are not architecturally dependent on (many) other, they would be more successful in implementing CDE and CD [S57]. The work reported in [S7] has conducted an empirical study on three projects that had adopted CI and CDE. The study concluded that most of the decisions (e.g., removing web services and collapsing the middle tier) made to achieve the desirable state of deployment (i.e., deployability quality attribute) were architectural ones. The collected deployability goals and tactics from three projects have been used as building blocks for the deployability tactics tree. Two studies [S5, S30] recommend that the component interfaces be clearly defined for making continuous delivery- or deployment-ready architectures.

**TABLE 9.** A classification of practices and lessons learnt for successfully implementing CI, CDE, and CD

| | | Practices | Key Points and Included Papers | # |
|---|---|---|---|---|
| Common Practices for Implementing CI, CDE, CD | Team Awareness and Communication | **PR1**. Improve Team Awareness and Communication | ▪ Listing the changed features in changelog entries [S43]<br>▪ Labelling the latest version and new features [S6, S47]<br>▪ Informing team members about branches that are completely out-dated [S49]<br>▪ Improved knowledge sharing between technical and management staffs on different levels [S6, S31, S37, S44] | 7 |
| | Investment | **PR2**. Planning and documentation | ▪ A planned path for adopting continuous practices [S28, S31, S36, S57, S58]<br>▪ Document builds, tests and other activities in integration processes [S6, S11, S17, S37, S43]<br>▪ Integration Flow Model [S11] | 10 |
| | | **PR3**. Promote team mindset | ▪ Organizing events about continuous practices to spread mindset and train team members [S5, S6, S45, S48]<br>▪ Giving much freedom to developers [S57]<br>▪ Empowering culture [S6, S57] | 5 |
| | | **PR4**. Improve team qualification and expertise | ▪ Formal training and coaching team members [S5, S6, S45, S48, S57] | 5 |
| | Clarifying Work Structures | **PR5**. Define new roles and teams | [S1, S9, S29, S30, S37, S45, S48, S49, S51, S57, S58]<br>▪ Establishing a dedicated team to develop and maintain deployment pipeline [S57, S58]<br>▪ Sheriff engineer [S29]<br>▪ Piloting team [S5]<br>▪ Virtual Scrum team [S37] | 11 |
| | | **PR6**. Adopt new rules and policies | [S26, S39, S45, S46, S48, S50, S58]<br>▪ All developers should be on call when releasing software [S58]. | 7 |
| Practices for Implementing CI | Improve Testing Activity | **PR7**. Improve Testing Activity | ▪ Practicing test-driven development [S12, S50]<br>▪ Test Planning practice [S17]<br>▪ Cross-team testing practice [S39]<br>▪ Designing decoupled tests by separating unit tests from functional and acceptance tests [S50] | 4 |
| | Branching Strategies | **PR8**. Branching Strategies | ▪ Using integration or local repository [S29, S48]<br>▪ Short-lived feature branching [S43]<br>▪ Practice of repository use [S30, S44]<br>▪ Not too many branches [S48] | 5 |
| | Decompose Development into Smaller Units | **PR9.** Decompose Development into Smaller Units | [S5, S10, S30, S36, S45, S47, S48, S49, S50, S51, S57]<br>▪ Dead code practice [S5]<br>▪ Breaking down large features and changes into smaller and safer ones [S10, S48, S49]<br>▪ Small and independent teams [S50, S51, S57] | 11 |
| Practices for Implementing CDE | Flexible and Modular Architecture | **PR10.** Flexible and Modular Architecture | [S5, S7, S12, S28, S30, S45, S51, S57]<br>▪ Deployability concern in mind when designing software systems [S7]<br>▪ Defining component interface clearly [S5, S30] | 8 |
| | Engage all people in Deployment | **PR11**. Engage all people in deployment process | ▪ Developer and tester take more responsibility about their code [S6, S9, S43, S44, S48, S57, S58]<br>▪ On call developers [S48] | 7 |
| Practices for Implementing CD | Partial Release | **PR12**. Partial Release | ▪ Zero release (Empty release) [S1]<br>▪ Hiding and disabling new or problematic functionalities to users [S6, S17, S44, S48]<br>▪ Deploying software to small set of users [S17, S44, S57]<br>▪ Rolling back quickly to stable state [S48]<br>▪ Independent releases [S58] | 7 |
| | Customer Involvement | **PR13**. Customer Involvement | [S10, S12, S28, S36, S43, S44, S49, S61, S63]<br>▪ Lead customer [S10, S12]<br>▪ Pilot customer [S43]<br>▪ Involving customers in testing phase [S61, S63]<br>▪ Triage meeting [S36] | 9 |

#### *b) Engage all people in deployment process*

A set of papers [S6, S9, S43, S44, S48, S57, S58] argued that achieving real benefits of continuous delivery and deployment practices requires developers and testers being more responsible for their codes in production environment. With this new responsibility, they are involved in and aware of all the steps (e.g., deploy into production), and are forced to fix problems that appear after deployment [S44]. As an example of involving developers in release process, Facebook adopted a policy, in which all engineers team who committed code should be on call during the release period [S48].

### *4) Practices for implementing CD*

#### *a) Partial Release*

Releasing software to customers potentially may be risky for software providers as their customers may receive buggy software. This issue can intensify when deploying software on a continuous basis (i.e., practicing CD). It is critical for software organizations to adopt practices in order to reduce potential risks and issues in release time. We identified three types of practices for this purpose: (i) deploying software to small set of users [S44, S17, S57]; (ii) hiding and disabling new or problematic functionalities to users [S6, S17, S44, S48]; (iii) rolling back quickly to stable state [S48]. Three papers [S17, S44, S48] pointed out *dark* and *canary* deployment methods that can significantly help transit to continuous deployment. In canary deployment method, the new versions of software are incrementally deployed to production environment with only a small set of users affected [49]. Deploying software by this method enables team to understand how new code (i.e., the canary) works compared to the old code (i.e., the baseline). In [S57], it was found that both Facebook and OANDA released software products to a small subset of users rather than releasing them to all customers. For example, Facebook first releases the software products to its own employees to get feedback to improve the test coverage. Another incremental release method, dark deployment, hides the functional aspects of new versions to end-users [50]. This method tries to detect potential problems, which may be caused by new versions of software before end-users would be affected. In order to deal with the large features (i.e., dark features) in OnDemand software product that may not be developed and deployed in a small cycle, one organization [S6] employed the practice of small batches. Through this practice, the development process of dark features was hidden from customers. However, when the entire feature is finally developed, the switch of dark feature will be turned on and then customer is able to interact with and use them. Another study [S58] reported the implementation of microservices that were independently released while maintaining backward compatibility with each release as a tactic of addressing delays in deployment pipeline. In order to introduce CD practice to novice developers, Krusche and Alperowitz [S1] suggested "***empty release***" practice, in which besides development teams get in touch with continuous workflows and infrastructures from day 0, continuous pipeline is initially run with simple application (e.g., "hello world").

#### *b) Customer Involvement*

Several papers [S10, S12, S28, S36, S43, S44, S49, S61, S63] aimed at exploring the role of customers or end-users as enabler in transition towards continuous deployment. A couple of papers [S10, S12] defined the concept of "***lead customer***", at which customers not only are incorporated in software development process, but also are eager to explore the concept of continuous deployment. The work reported in [S43] used the term "***pilot customer***" and argued that it would be better to apply CDE or CD to those companies that are willing to continuously receive updates. It has been noted that it is needed to renew existing engagement model with customers to be compatible with the spirit of CD. Agarwal [S36] described a process model based on Type C SCRUM, called Continuous SCRUM, and leveraged a number of best practices to augment this process model and achieve sustainable weekly release. One of the noticeable practices was "***triage meeting***", in which product-owner runs the meeting and she/he determines the triage committee. A product-owner review has been introduced into the sprint to enable and approve changes to product requirements as well as the product-owner was enabled to prioritize the back-log of product requirements. We found a set of papers [S61, S63] arguing the involvement of customer in testing was an effective practice for adopting CDE and CD practices. A study [S61] revealed that involving customers in testing phase is a helpful practice for those companies that do not have enough resources for practicing CD. The study indicated that customers can be greatly successful in finding lower impact functional defects.

# V. Discussion

Continuous practices (i.e., Continuous Integration (CI), Continuous DElivery (CDE), and Continuous Deployment (CD)) are increasingly becoming popular in software industry. Several dozens of approaches, tools, challenges, and practices have been reported for adopting, implementing and promoting CI, CDE, and CD. It is equally important to systematically review and thoroughly document the reported approaches, tools, challenges, and practices as a body of knowledge. Such body of knowledge can help understand their nature and potential areas of applications and identify the areas of future research direction. The abovementioned needs stimulated four key research questions to be answered through this SLR. The previous section has presented the findings from this SLR with respect to the research questions. Now we discuss the findings and reflect upon the potential areas for further research.

## A. Mapping of Challenges to Practices

Figure 5 presents a mapping of the identified challenges in Section IV.D onto the practices reported in Section IV.E. This mapping is intended to provide a reader (i.e., researcher or practitioner) to quickly determine which challenges are related to which practices. For example, a *flexible and modular architecture* is expected to decrease *dependencies in design and code*. Figure 5 also indicates that that there might be dependencies among the challenges (i.e., *exacerbation*) or practices (i.e., *support*). A practice may support or positively affect another practice, for example, by making the implementation of that practice easier. For example, we found that *distributed organization* can exacerbate the challenge of and need for *coordination and collaboration* in adopting continuous practices; however, adopting and implementing *partial release* can be greatly supported by *engaging all people (in particular customer) in deployment process*.

## B. Critical factors for continuous practices success

Based on our analysis in Sections IV.D and IV.E, we have identified 20 challenges and 13 practices for CI, CDE, and CD. We have also found 30 approaches and associated tools that have been proposed by the reviewed studies to address particular challenges in each continuous practice. It is important to point out that there was no one-to-one relationship between the identified challenges and the proposed practices, approaches and associated tools as there were some challenges for which we were unable to identify any practice or approaches to address them and vice versa. We decided to define a set of critical factors that should be carefully considered to make continuous practices successful. To identify what factors (i.e., both in software development and customer organizations) are important to successfully adopt and implement continuous practices, we again analyzed the results reported in Sections IV.B, IV.D, and IV.E. A factor is accumulated challenges, approaches, and practices pertaining to a fact. For example, we found a number of challenges (Sections IV.D.2.a), approaches and associated tools (Sections IV.B.1 and IV.B.3), and practices (Section IV.E.2.a) for testing activity in moving towards continuous practices. Therefore, we considered "testing" as a factor, which should be carefully considered when adopting continuous practices. If a factor is cited in at least 20% of the reviewed studies then we regard that factor as a critical factor for making continuous practices successful.

Table 10 shows the list of 7 critical factors, which may impact the success of continuous practices. "Testing" (27 papers, 39.1%) is the most frequently mentioned factor for continuous practices success, followed by "team awareness and transparency" (24 papers, 34.7%), "good design principles" (21 papers, 30.4%) and "customer" (17 papers, 24.6%). Our results indicate that "testing" plays an important role in successfully establishing continuous practices in a given organization. Our research reveals that long running tests, manual tests, and high frequency of test cases failure have failed most of the case organizations in the reviewed studies to realise and achieve the anticipated benefits of continuous practices. Whilst we have reviewed several papers that revealed a lack of test automation was a roadblock to move toward continuous practices, there were only a few papers (i.e., 7 papers), which had developed and proposed approaches, tools and practices for automating tests for this purpose.

Continuous practices promise to significantly reduce integration and deployment problems. It should be designed in a way that the status of a project, number of errors, who broke the build, and the time when features are finished are visible and transparent to all team members. We have found "team awareness and transparency" as the second-most critical factor for adopting continuous practices. Improved team awareness and transparency across the entire software development enables team members to timely find potential conflicts before delivering software to customers and also improves collaboration among all teams [51].

Our review has identified 17 papers that report challenges, practices and lessons learnt regarding customers, which enabled us to consider "customer" as a critical factor for successful implementation of continuous practices. It is worth mentioning that this factor mostly impacts on CD. We found that not always customer organizations are happy with continuous release. That is why we need to investigate the level of customer satisfaction when moving to CD practice: unavailability of customer environments, extra computing resources required from customers, incompatibility of new release with existing components and systems, and increased chance of receiving buggy software all together can demotivate customers about advantages of continuous deployment. Our results also indicate that "highly skilled and motivated team" (15 out of 69, 21.7%) is a critical factor to drive software organizations towards continuous practices. We argue that releasing continuously and automatically software can be achieved with solid foundation of technical and soft skills, shared responsibilities among team members, and having motivated teams to continuously learn new tools and technologies.

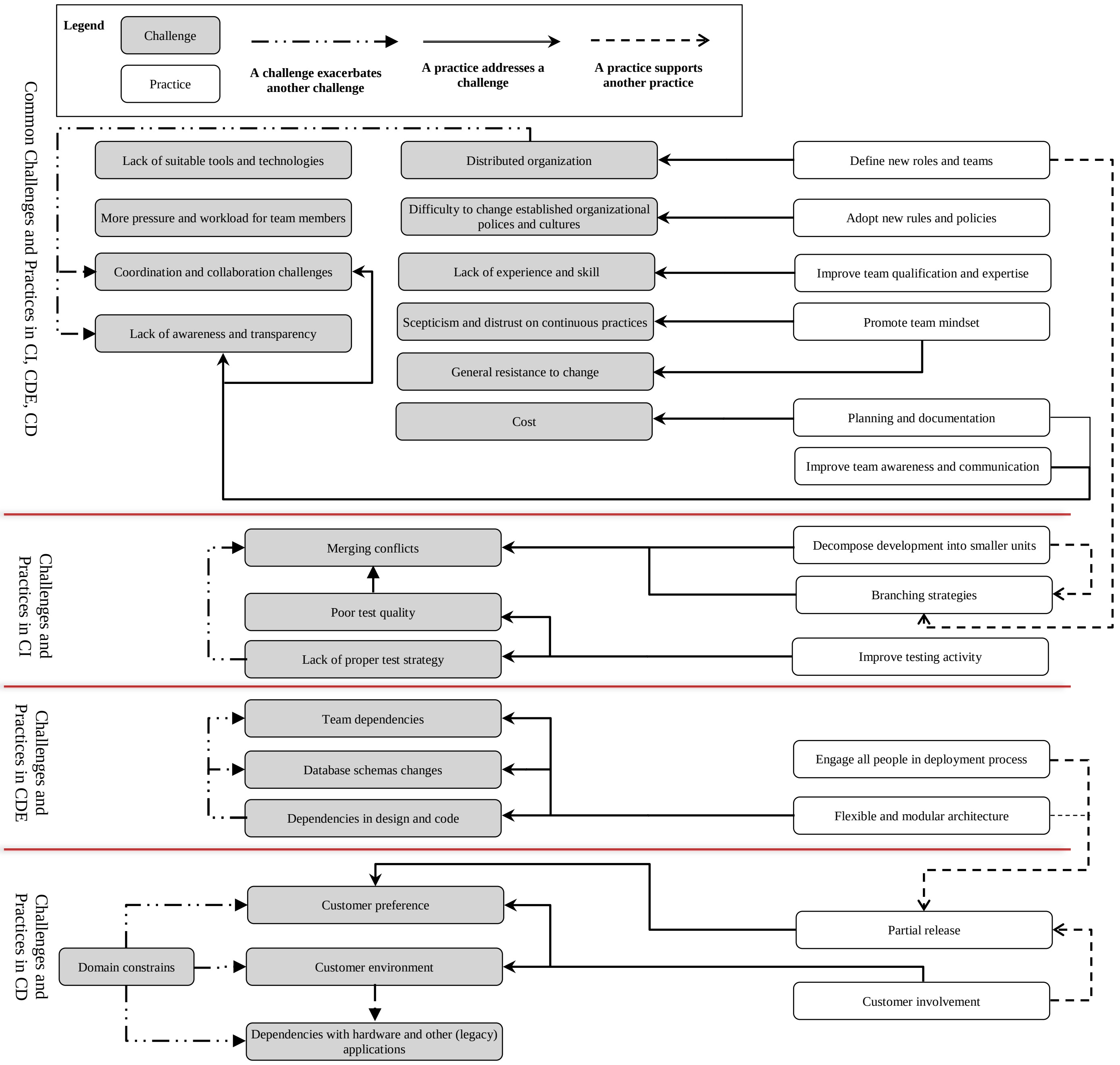

**FIGURE 5**. An overview of challenges and practices of adopting CI, CDE, CD, and the relationship among them

**TABLE 10**. List of critical factors for continuous practices success

| ID | Factor | # | % | Studies |
|---|---|---|---|---|
| **F1** | Testing (effort and time) | **27** | **39.1** | S3, S4, S5, S6, S12, S17, S19, S23, S25, S32, S34, S36, S38, S39, S40, S41, S43, S45, S50, S52, S53, S54, S55, S56, S62, S64, S67 |
| **F2** | Team awareness and transparency | **24** | **34.7** | S1, S2, S4, S6, S10, S13, S22, S24, S31, S33, S37, S38, S41, S43, S44, S45, S47, S49, S50, S52, S56, S62, S64, S67 |
| **F3** | Good design principles | **21** | **30.4** | S4, S5, S6, S7, S12, S10, S28, S30, S31, S36, S41, S45, S47, S48, S49, S50, S51, S57, S58, S60, S62 |
| **F4** | Customer | **17** | **24.6** | S4, S6, S10, S12, S28, S29, S36, S43, S44, S49, S55, S56, S57, S60, S61, S62, S63 |
| **F5** | Highly skilled and motivated team | **15** | **21.7** | S4, S5, S6, S9, S12, S43, S44, S45, S48, S49, S57, S56, S57, S58, S62 |
| **F6** | Application domain | **14** | **20.2** | S4, S5, S6, S9, S10, S24, S31, S41, S44, S48, S56, S57, S60, S65 |
| **F7** | Appropriate infrastructure | **14** | **20.2** | S5, S6, S8, S10, S27, S43, S47, S49, S56, S57, S59, S60, S66, S68 |

Whilst this SLR reveals that continuous practices have been applied successfully to both maintenance and greenfield projects, we argue that "application domain" can play a significant role in transition towards continuous practices, in particular continuous deployment. As discussed earlier, continuous delivery can be applied to all types of applications and organizations. However, practicing CD in some application domains (e.g., embedded systems domain) is associated with unique challenges, in which they make almost impossible to truly practice CD or affect the frequency of releases to customer environments. We emphasize that application domains and limitations of customers should be carefully studied before adopting continuous deployment. Our SLR reveals that one of the leading causes of failure in fully implementing continuous practices is missing or poor infrastructures. By "appropriate infrastructure", we mean all software development tools, infrastructures, networks, technologies and physical resources (e.g., build server and test automation servers) employed by an organization to do continuous practices well. This is mainly because implementing each continuous practice, in particular continuous delivery and deployment in a given organization requires extra computing resources and also tools and technologies to automate end-to-end software development (e.g., testing) and release process as much as possible. This consequently would affect organizational budget. We assert that one of the core components of an appropriate infrastructure, which considerably enables automation support and impact the success of continuous practices, is deployment pipeline. We will concretely discuss the engineering process of deployment pipeline in Section V.E.

## *C. Contextual factor*

The importance of contextual attributes and what should be reported as contextual attributes have been discussed in the software engineering literature [52-54]. It has been argued that software development approaches, tools, challenges, lessons learnt and best practices need to be explored and understood along with their respective contexts [53, 55]. Particularly, we tried to understand in which methodological and organizational contextual settings (i.e., research type, project type, application domain, organization size and domain) the proposed approaches, tools, best practices and challenges have been reported. According to the results reported in Section IV.A.2, the reviewed studies were evaluation research (25 papers, 36.2%), followed by validation research (24 papers, 34.7%) and experience report (15 papers, 21.7%). Since all of the experience papers were based on practitioners' experiences, the combination of both evaluation and experience papers means that 57.9% of the reviewed papers came from industry setting. The high percentage of the papers with industrial level evidence improves the practical applicability of the reported results and encourages practitioners to adopt and employ the proposed approaches, tools, practices and consider the challenges when adopting each continuous practice. As reported in Section IV.A.4, a considerable number of the reviewed papers did not provide the information on application domain and type, resulting in these papers being categorized as "unclear". There was a general lack of information about the organizational contexts (i.e., size and domain) in the reviewed papers. We were forced to drop them for data analysis and interpretation. We strongly suggest that more attention be paid to reporting the contextual information about the reported studies. The contextual information is likely to improve the quality and credibility of the reported approaches, tools and practices in continuous integration, delivery and deployment. Such information can also help a reader to better understand the reported research.

## *D. Architecting for deployability*

The results of this SLR indicate that sound architecture design (i.e., "good design principles" factor) has a significant influence on the success of practicing CI, CDE, and CD. Several of the reviewed papers have discussed modular architecture, loosely coupled components, and clearly defined interfaces as contributing factors for adopting and implementing continuous practices, in particular CDE and CD. Based on Section IV.D.4.a, the importance of this issue increases sharply in heterogeneous environments that can hinder continuous software deployment. We argue that one of the most pressing challenges of adopting and implementing continuous practices is how software applications should be (re-) architected to develop, integrate, test and deploy independently in multiple environments. Therefore, the architecting phase should be considered as one of the most important phases for appropriately adopting and implementing continuous practices [56]. Deployability as an emerging quality attribute has a high priority for continuous delivery and deployment [15, 57, 58]. By deployability, we mean *"how reliably and easily an application/component/service can be deployed to (heterogeneous) production environment"* [58].

Architecting with testability and deployability in mind during the design time has been featured in many white papers and practitioners' blogs [15, 57] as a noticeable practice for CDE and CD, but we could find only one paper [S7] that has explicitly considered the deployability scenarios for upfront design decisions and concluded that most of the decisions made for deployment-related issues were architectural one. We assert that there is an important need of research to gain a deep understanding of how continuous delivery or deployment adoption can influence the architecting process and their outcomes in an organisation. We argue that this research area (i.e., architecting for deployability) should be more investigated in the future. This motivates the following questions: How can we evaluate and measure the deployability of a designed architecture at the early stage of development time? What quality attributes are in support of or in conflict with deployability? Which architectural patterns, tactics, and styles are more-friendly for deployability?

## *E. Engineering deployment pipeline*

In Section IV.C, we have discussed that deployment pipeline is a key enabler for enterprises to successfully adopt continuous practices. Our review has revealed that despite a significant number of the reviewed papers conducted in industrial settings and reported by practitioners, many papers lacked sufficient details about how enterprises design and implement deployment pipelines and what challenges they might experience. In fact, only 36.2% of the included studies presented the tools, which have been employed to implement deployment pipelines. This investigation was interesting because there is no standard or single pipeline [1] and modelling and implementing a deployment pipeline in a given enterprise may be influenced by a number of factors such as team skills, experience and structure, organization's structure and budget, customer environments, and project domain [43]. Therefore, software development organizations need to allocate time and resources to appropriately select and integrate a wide variety of open source and commercial tools to form a deployment pipeline tailored to them. The evidence of this growing need is the emergent of consulting companies such as Sourced Group[14] and Xebia[15] that are assisting enterprises in designing and implementing deployment pipeline.

In the meanwhile, with the increasing size and complexity of software-intensive systems, the number of builds and test cases increase dramatically. Whilst infrastructures with high-performance computing resources and selecting appropriate tools are mandatory for implementing continuous practices and deployment pipeline, this is not sufficient to deal with such tremendous growth rate. Therefore, it is needed to develop innovative approaches and tools, which not only enable team members to receive build and test results correctly and timely, but also they should be aligned and integrated with deployment pipeline. In Section IV.B, thirty approaches and associated tools have been reported to support and facilitate continuous practices. Most of them (24 out of 30) only target CI practice; 18 out of 30 are stand-alone tools that have not been integrated and evaluated in a deployment pipeline. Another increasing concern in the deployment pipeline is how to secure a deployment pipeline [59]. According to [59], the main concern raised during RELENG[16] workshop in 2014 was "what happens if someone subverts the deployment pipeline". All stages and tools involved in the deployment pipeline as well as integrating application to other infrastructures can potentially be compromised by attackers. Two papers [S27, S66] have investigated the security issue in deployment pipelines. We conclude that there is a paucity of research aimed at systematically studying engineering process of deployment pipelines. We assert software engineering researchers and practitioners need to pay more attention to systematically architect deployment pipelines and rigorously selecting appropriate tools for the pipelines.

## VI. THREATS TO VALIDITY

Whilst we strictly followed the guidelines provided by [25], we had similar validity threats like other SLRs in software engineering. The findings of this SLR may have been affected by the following threats:

**Search strategy**: One of the threats that may occur in any SLR is the possibility of missing or excluding the relevant papers. To mitigate this threat, as discussed in Section III.B.3, we used six popular digital libraries to retrieve the relevant papers. We argue that using Scopus as the largest indexing system which provides the most comprehensive search engine among other digital libraries [55], enabled us to increase the coverage of the relevant studies. Additionally, we employed three strategies to mitigate any potential threat in the search strategy: i) search string was improved iteratively based on the pilot search and were tested carefully before executing for searching the relevant papers for this review; 2) we consulted the search strings used in the existing SLRs [12, 13] for building our search string; 3) a snowballing technique (i.e., manual search on references of the selected papers) was employed in the second round of the papers searching process (see Figure 2) to identify as many related papers as possible.

**Study selection:** This step can be influenced by researchers' subjective judgement about whether or not a paper meets the selection criteria for inclusion or exclusion. The potential biases in the study selection have been addressed by strictly following the pre-defined review protocol, recording the inclusion and exclusion reasons for on-going internal discussions among first and second authors about the papers that raised doubts about their inclusion or exclusion decisions. At the first step, the inclusion and exclusion criteria have been validated by the first two authors on a small subset of primary studies. Any disagreements during study selection were resolved through discussions between them. Furthermore, the second and third authors performed a cross-check using a random number of the selected papers.

**Data extraction**: Researchers' bias in data extraction can be a basic threat in any SLR, which may negatively affect the results of SLRs. We implemented the following steps to address this threat. First we created a data extraction form (see Table 12) to consistently extract and analyze the data for answering the research questions of this SLR. Second, since a large part of the data extraction step was conducted by the first author; in the case of any doubt, continuous discussions were organised with the second author for correcting any disparities in the extracted data. Third, a subset of the extracted data was verified by the second and third authors.

**Data synthesis**: As we argued in Section III.E.2, we applied quantitative and qualitative methods to analyze the extracted data. It should be noted that sometime there were some difficulties in interpreting the extracted data due to lack of sufficient information about the data items. We had to subjectively interpret and analyze the data items, which might have had an effect on the data extraction outcomes. To reduce the researchers' bias in interpretation of the results, besides reading the given study, where possible we also referred the approach's and tool's website and any training movie (e.g., RQ1 and RQ2) to get more reliable information. It should be noted that for other data items, we did not have any interpretation unless the data items have been explicitly provided by the study (e.g., application domain).

## VII. CONCLUSIONS AND IMPLICATIONS

This work has presented a Systematic Literature Review (SLR) of approaches, tools, challenges and practices identified in empirical studies on continuous practices in order to provide an evidential body of knowledge about the state of the art of continuous practices and the potential areas of research. We selected 69 papers from 2004 to 1st June 2016 for data extraction, analysis, and synthesis based on pre-defined inclusion and exclusion criteria. A rigorous analysis and systematic synthesis of the data extracted from the 69 papers have enabled us to conclude:

(1) The research on continuous practices, in particular continuous delivery and deployment is gaining increasing interest and attention from software engineering researchers and practitioners according to the steady upward trend in the number of papers on continuous practices in the last decade (see Figure 3). More than half

[14] http://www.sourcedgroup.com/
[15] https://xebia.com/
[16] http://releng.polymtl.ca/RELENG2014/html/

of the reviewed papers (39 papers, 56.5%) have been published in the last three years.

(2) With respect to the research type, most of the selected papers were evaluation (25 out of 69, 36.2%) and validation (24 out of 69, 34.7%) research papers. While 21.7% of the selected papers were experience papers, a small number of papers were solution proposal (7.2%). A large majority of the papers were conducted in industrial (i.e., 64 out of 69, 92.7%) rather than academic (i.e., 5 papers) settings. With respect to the data analysis approach, the same number of the selected papers used quantitative and qualitative research approaches (i.e., 37.6% for each), while this statistic was 20.2% for mixed approaches.

(3) The approaches, tools, challenges, and practices reported for adopting and implementing continuous practices have been applied to a wide range of application domains, and among which ‘‘software/web development framework’’ and “utility software” have received the most attention. This SLR also revealed that continuous practices can be successfully applied to both greenfield and maintenance projects.

(4) Thirty approaches and associated tools have been identified by this SLR, which facilitate the implementation of continuous practices in the following ways (i.e., not mutually exclusive): *reducing build and test time in CI* (9 approaches), *increasing visibility and awareness on build and test results in CI* (10 approaches), *supporting (semi-) automated continuous testing* (7 approaches), *detecting violations, flaws and faults in CI* (11 approaches), *addressing security, scalability issues in deployment pipeline* (3 approaches), and *improve dependability and reliability of deployment process* (3 approaches).

(5) We observed that only 36.2% of the selected papers reported what and how tools and technologies were selected and integrated to implement deployment pipeline (i.e., modern release pipeline). *Subversion* and *Git/GitHub* as version control systems and *Jenkins* as integration server were the most popular tools used in deployment pipelines.

(6) The identified approaches (see Section IV.B), challenges (see Section IV.D) and practices (see Section IV.E) of CI, CDE, and CD have enabled us to find seven critical factors that impact the success of continuous practices, in an order of importance: “*testing (effort and time)*”, “*team awareness and transparency*”, “*good design principles*”, “*customer*”, “*highly skilled and motivated team*”, “*application domain*”, and “*appropriate infrastructure*”.

(7) Implications for researchers: (i) this SLR has revealed the scarcity of reporting contextual information (e.g., organization size and domain) in the selected papers. To improve the quality and credibility of the results, researchers ought to report detailed contextual information. (ii) In this review, we found only two papers that investigated the security issue in deployment pipelines. Given the increased importance of security in deployment pipelines, there is a need of further research to explore how deployment pipelines should be designed and implemented to mitigate security issues. (iii) Out of 30 approaches and associated tools reported in this SLR, only 12 approaches and tools were integrated and evaluated in deployment pipeline. We encourage researchers to evaluate their proposed approaches and tools with real deployment pipelines. (v) As discussed in Section V.D, architecture design and deployability quality attribute are very important factors in successfully adopting and implementing continuous practices, however, there is a lack of guidance of architecting for deployability. We suggest that researchers in cooperation with practitioners come up with frameworks, processes, and tools to support deployability quality attribute at design time.

(8) Implications for practitioners: (i) a very high percentage of the reviewed papers provide industrial level evidence (i.e., evaluation and practitioners’ experience papers as presented in Section IV.A.2). This improves the practical applicability of the reported results. Such findings are expected to encourage software engineering practitioners to adopt and employ appropriate approaches, tools, practices and consider the reported challenges in their daily work based on the suitability for different contexts. (ii) The identified approaches, tools, challenges, and practices have been classified in a way that practitioners are enable to understand what challenges are for adopting each continuous practice, what approaches and practices exist for supporting and facilitating each continuous practice. We found a number of challenges and practices that were common in transition towards all CI, CDE, and CD. (iii) The identified critical factors can make practitioners aware of the factors that may affect the success of continuous practices in their organizations. For example, whilst it is important for practitioners to know that a lack of team awareness and transparency may fail them to realise and achieve the real anticipated benefits of continuous practices, this SLR has identified several approaches, associated tools and practical solutions to improve and sustain team awareness and transparency in continuous practices.

## ACKNOWLEDGMENT

This work is partially supported by Data61, a business unit of CSIRO, Australia. The first author is also supported by Australian Government Research Training Program Scholarship.

## Appendix A. Selected studies

**TABLE 11.** Selected studies in the review

| ID | Title | Author(s) | Venue | Year |
|---|---|---|---|---|
| S1 | Introduction of continuous delivery in multi-customer project courses | S. Krusche, L. Alperowitz | International Conference on Software Engineering | 2014 |
| S2 | SQA-Mashup: A mashup framework for continuous integration | M. Brandtner, E. Giger, H. Gall | Information and Software Technology | 2015 |
| S3 | Vroom: Faster build processes for java | J. Bell, E. Melski, M. Dattatreya, G.E. Kaiser | IEEE Software | 2015 |
| S4 | The highways and country roads to continuous deployment | M. Leppänen, S. Mäkinen, M. Pagels, V. Eloranta, J. Itkonen, M.V. Mäntylä, T. Männistö | IEEE Software | 2015 |
| S5 | Challenges when adopting continuous integration: A case study | A. Debbiche, M. Diener, R.B. Svensson | International Conference on Product-Focused Software Process Improvement | 2014 |
| S6 | On the journey to continuous deployment: Technical and social challenges along the way | G. Claps, R.B. Svensson, A. Aurum | Information and Software Technology | 2015 |
| S7 | Toward design decisions to enable deployability: Empirical study of three projects reaching for the continuous delivery holy grail | S. Bellomo, N. Ernst, R. Nord, R. Kazman | International Conference on Dependable Systems and Networks | 2014 |
| S8 | Achieving reliable high-frequency releases in cloud environments | L. Zhu, D. Xu, A.B. Tran, X. Xu, L. Bass, I. Weber, S. Dwarakanathan | IEEE Software | 2015 |
| S9 | The practice and future of release engineering: A roundtable with three release engineers | B. Adams, S. Bellomo, C. Bird, T. Marshall-Keim, F. Khomh, K. Moir | IEEE Software | 2015 |
| S10 | Climbing the "Stairway to heaven" - A mulitiple-case study exploring barriers in the transition from agile development towards continuous deployment of software | H.H. Olsson, H. Alahyari, J. Bosch | Euromicro Conference on Software Engineering and Advanced Applications | 2012 |
| S11 | Automated software integration flows in industry: A multiple-case study | D. Ståhl, J. Bosch | International Conference on Software Engineering | 2014 |
| S12 | Hitting the target: Practices for moving toward innovation experiment systems | T. Karvonen, L.E. Lwakatare, T. Sauvola, J. Bosch, H.H. Olsson, P. Kuvaja, M. Oivo | International Conference on Software Business | 2015 |
| S13 | Visualizing testing activities to support continuous integration: A multiple case study | A. Nilsson, J. Bosch, C. Berger | International Conference on Agile Software Development (XP) | 2014 |
| S14 | Implementation of continuous integration and automated testing in software development of smart grid scheduling support system | J. Lu, Z. Yang, J. Qian | International Conference on Power System Technology | 2014 |
| S15 | Implementing continuous integration software in an established computational chemistry software package | R.M. Betz, R.C. Walker | International Workshop on Software Engineering for Computational Science and Engineering | 2013 |
| S16 | Making software integration really continuous | M.L. Guimarães, A.R. Silva | International Conference Fundamental Approaches to Software Engineering | 2012 |
| S17 | Continuous delivery? Easy! Just change everything (well, maybe it is not that easy) | S. Neely, S. Stolt | Agile Conference (AGILE) | 2013 |
| S18 | Software product measurement and analysis in a continuous integration environment | G. de Souza Pereira Moreira, R.P. Mellado, D.Á. Montini | International Conference on Information Technology: New Generations | 2010 |
| S19 | UBuild: Automated testing and performance evaluation of embedded linux systems | F. Erculiani, L. Abeni, L. Palopoli | International Conference on Architecture of Computing Systems | 2014 |
| S20 | Using continuous integration of code and content to teach software engineering with limited resources | J.G. Süβ, W. Billingsley | International Conference on Software Engineering | 2012 |
| S21 | Backtracking incremental continuous integration | T. van der Storm | European Conference on Software Maintenance and Reengineering | 2008 |
| S22 | BuildBot: Robotic monitoring of agile software development teams | R. Ablett, E. Sharlin, F. Maurer, J. Denzinger, C. Schock | International Conference on Robot & Human Interactive Communication | 2007 |
| S23 | Mixed data-parallel scheduling for distributed continuous integration | O. Beaumont, N. Bonichon, L. Courtes, E. Dolstra, X. Hanin | International Parallel and Distributed Processing Symposium Workshops & PhD Forum | 2012 |
| S24 | SQA-Profiles: Rule-based activity profiles for Continuous Integration environments | M. Brandtner, S.C. Muller, P. Leitner, H.C. Gall | International Conference on Software Analysis, Evolution, and Reengineering | 2015 |
| S25 | Identifying and understanding header file hotspots in C/C++ build processes | S. McIntosh, B. Adams, M. Nagappan, A.E. Hassan | Automated Software Engineering | 2015 |
| S26 | Practical experience with test-driven development during commissioning of the multi-star AO system ARGOS | M. Kulas, J.L. Borelli, W. Gässler, D. Peter, S. Rabien, G.O. de Xivry, L. Busoni, M. Bonaglia, T. Mazzoni, G. Rahmer | Software and Cyberinfrastructure for Astronomy III | 2014 |
| S27 | Security of public continuous integration services | V.Gruhn, C. Hannebauer, C. John | International Symposium on Open Collaboration | 2013 |
| S28 | Elaboration on an integrated architecture and requirement practice: Prototyping with quality attribute focus | S. Bellomo, R. L. Nord, I. Ozkaya | International Workshop on the Twin Peaks of Requirements and Architecture | 2013 |
| S29 | Rapid releases and patch backouts: A software analytics approach | R. Souza, C. Chavez, R.A. Bittencourt | IEEE Software | 2015 |
| S30 | Patterns for continuous integration builds in cross-platform agile software development | C. Hsieh, C. Chen | Journal of Information Science and Engineering | 2015 |
| S31 | Technical dependency challenges in large-scale agile software development | N. Sekitoleko, F. Evbota, E. Knauss, A. Sandberg, M. Chaudron, H. H. Olsson | International Conference on Agile Software Development (XP) | 2014 |
| S32 | A technique for agile and automatic interaction testing for product lines | M.F. Johansen, Ø. Haugen, F. Fleurey, E. Carlson, J. Endresen, T. Wien | International Conference on Testing Software and Systems | 2012 |
| S33 | Ambient awareness of build status in collocated software teams | J. Downs, B. Plimmer, J. G. Hosking | International Conference on Software Engineering | 2012 |
| S34 | How well do test case prioritization techniques support statistical fault localization | B. Jiang, Z. Zhang, W.K. Chan, T.H. Tse, T.Y. Chen | Information and Software Technology | 2012 |
| S35 | Integrating early V&V support to a GSE tool integration platform | J.P. Pesola, H. Tanner, J. Eskeli, P. Parviainen, D. Bendas | International Conference on Global Software Engineering Workshops | 2011 |
| S36 | Continuous SCRUM: Agile management of SAAS products | P. Agarwal | India Software Engineering Conference | 2011 |
| S37 | Hitting the wall: What to do when high performing scrum teams overwhelm operations and infrastructure | J. Sutherland, R. Frohman | Hawaii International Conference on System Sciences | 2011 |
| S38 | Test automation framework for implementing continuous integration | E.H. Kim, J. Chae Na, S.M. Ryoo | International Conference on Information Technology: New Generations | 2009 |
| S39 | Using continuous integration and automated test techniques for a robust C4ISR system | H.M. Yüksel, E. Tüzün, E. Gelirli, B. Baykal | International Symposium on Computer and Information Sciences | 2009 |
| S40 | A Unified test framework for continuous integration testing of SOA solutions | H. Liu, Z. Li, J. Zhu, H. Tan, H. Huang | International Conference on Web Services | 2009 |
| S41 | Factors impacting rapid releases: An industrial case study | N. Kerzazi, F. Khomh | International Symposium on Empirical Software Engineering and Measurement | 2014 |
| S42 | Ultimate architecture enforcement custom checks enforced at code-commit time | P. Merson | Companion of Conference on Systems, Programming, & Applications: Software for Humanity | 2013 |
| S43 | Transitioning towards continuous delivery in the B2B domain: A case study | O. Rissanen, J. Münch | International Conference on Agile Software Development (XP) | 2015 |
| S44 | Synthesizing continuous deployment practices used in software development | A.A.U. Rahman, E. Helms, L. Williams, C. Parnin | Agile Conference (AGILE) | 2015 |
| S45 | Stakeholder perceptions of the adoption of continuous integration-A case study | E. Laukkanen, M. Paasivaara, T. Arvonen | Agile Conference (AGILE) | 2015 |
| S46 | Toward agile architecture: Insights from 15 years of ATAM data | S. Bellomo, I. Gorton, R. Kazman | IEEE Software | 2015 |
| S47 | Enterprise continuous integration using binary dependencies | M. Roberts | International Conference on Agile Software Development (XP) | 2004 |
| S48 | Development and deployment at Facebook | D. G. Feitelson, E. Frachtenberg, K. L. Beck | IEEE Internet Computing | 2013 |
| S49 | Transforming a six month release cycle to continuous flow | M. Marschall | Agile Conference (AGILE) | 2007 |
| S50 | Scaling continuous integration | R.O. Rogers | International Conference on Agile Software Development (XP) | 2004 |
| S51 | Architectural tactics to support rapid and agile stability | F. Bachmann, R.L. Nord, I. Ozkaya | CrossTalk: The Journal of Defense Software Engineering | 2012 |
| S52 | Continuous automated testing of sdr software | J. Nimmer, B. Fallik, N. Martin, J. Chapin | Software Defined Radio Technical Conference | 2006 |
| S53 | Surrogate: A simulation apparatus for continuous integration testing in service oriented architecture | H.Y. Huang, H.H. Liu, Z.J. Li, J. Zhu | International Conference on Services Computing | 2008 |
| S54 | CiCUTS: Combining system execution modeling tools with continuous integration environments | J. H. Hill, D.C. Schmidt, A.A. Porter, J.M. Slaby | International Conference and Workshop on the Engineering of Computer Based Systems | 2008 |
| S55 | Techniques for improving regression testing in continuous integration development environments | S. Elbaum, G. Rothermel, J. Peni | International Symposium on Foundations of Software Engineering | 2014 |
| S56 | Towards DevOps in the Embedded Systems Domain: Why is It so Hard? | L. E. Lwakatare, T. Karvonen, T. Sauvola, P. Kuvaja, H. H. Olsson, J. Bosch, M Oivo | Hawaii International Conference on System Sciences | 2016 |
| S57 | Continuous deployment at Facebook and OANDA | T. Savor, M. Douglas, M. Gentili, L. Williams, K. Beck, M. Stumm | International Conference on Software Engineering | 2016 |
| S58 | DevOps making it easy to do the right thing | M. Callanan, A. Spillane | IEEE Software | 2016 |
| S59 | Rondo: A tool suite for continuous deployment in dynamic environments | O. Günalp, C. Escoffier, P. Lalanda | International Conference on Services Computing | 2015 |
| S60 | DevOps: A definition and perceived adoption impediments | J. Smeds, K. Nybom, I. Porres | International Conference on Agile Software Development (XP) | 2015 |
| S61 | Social Testing: A framework to support adoption of continuous delivery by small medium enterprises | J. Dunne, D. Malone, J. Flood | International Conference on Computer Science, Computer Engineering, and Social Media | 2015 |
| S62 | Automated testing in the continuous delivery pipeline: A case study of an online company | J. Gmeiner, R. Ramler, J. Haslinger | User Symposium on Software Quality, Test and Innovation | 2015 |
| S63 | Towards post-agile development practices through productized development infrastructure | M. Leppänen, T. Kilamo, T. Mikkonen | International Workshop on Rapid Continuous Software Engineering | 2015 |
| S64 | Supporting continuous integration by code-churn based test selection | E. Knauss, M. Staron, W. Meding, O. Söder, A. Nilsson, M. Castell | International Workshop on Rapid Continuous Software Engineering | 2015 |
| S65 | Requirements to pervasive system continuous deployment | C. Escoffier, O. Günalp, P. Lalanda | International Conference on Service-Oriented Computing Workshops | 2013 |
| S66 | Composing patterns to construct secure systems | P. Rimba, L. Zhu, L. Bass, I. Kuz, S. Reeves | European Dependable Computing Conference | 2015 |
| S67 | Fast feedback from automated tests executed with the product build | M. Eyl, C. Reichmann, K. Müller-Glaser | International Conference Software Quality Days | 2016 |
| S68 | POD-Diagnosis: Error diagnosis of sporadic operations on cloud applications | X. Xu, L. Zhu, I. Weber, L. Bass, D. Sun | International Conference on Dependable Systems and Networks | 2014 |
| S69 | Feature toggles: practitioner practices and a case study | M. T. Rahman, L. P. Querel, P. C. Rigby, B. Adams | Working Conference on Mining Software Repositories | 2016 |

# Appendix B. Data Extraction Form

**TABLE 12.** Data items extracted from each study and related research questions

| # | Data item | Description | RQs (Section III.A) |
|---|---|---|---|
| **D1** | Author(s) | The author(s) of the paper. | |
| **D2** | Year | The year of the publication of the paper | Demographic data |
| **D3** | Title | The title of the paper | |
| **D4** | Publication type | The type of publication (e.g., journal paper) | Demographic data |
| **D5** | Venue | The name of the publication venue | Demographic data |
| **D6** | Data analysis type | Qualitative, quantitative or mixed. | Demographic data |
| **D7** | Research type | The type of research i.e., validation research, evaluation research, solution proposal, philosophical paper, opinion paper, and experience report. | Demographic data |
| **D8** | Study context | The study contexts are categorized in industry and non-industry (e.g. student) cases | Demographic data |
| **D9** | Project type | It recodes the type of project e.g., greenfield or maintenance. | Demographic data |
| **D10** | Application domain | The type of application used for reporting challenges as well as for validating proposed techniques, tools, and practices. | Demographic data |
| **D11** | Techniques and tools | The techniques and tools that facilitate the continuous integration, delivery and deployment (i.e., continuous practices). | RQ1, RQ2 |
| **D12** | Challenges | It documents the challenges and barriers that have been reported to adopt continuous practices in software development and customer's organizations. | RQ3 |
| **D13** | Practices | It records lessons learned, authors' experiences and good practices to successfully implement continuous practices. | RQ4 |
| **D14** | Critical factors | Factors to be considered when introducing and adopting continuous practices. | Discussion |